\documentclass[12pt]{article}
\usepackage[USenglish]{babel}
\usepackage[a4paper, left=2cm, right=2cm, bottom=3cm, heightrounded]{geometry}

\usepackage{enumitem}
\usepackage{amsfonts}
\usepackage{amsmath}
\usepackage{mathrsfs}
\usepackage{mathabx}
\usepackage{amssymb}
\usepackage{nccmath}
\usepackage{upgreek}
\usepackage{bbm}
\usepackage{graphicx}
\usepackage{cancel}
\usepackage{color}
\usepackage{slashed}
\usepackage{cite}
\usepackage{hyperref}

\usepackage{booktabs}
\usepackage[dvipsnames]{xcolor}

\definecolor{Managed-Periwinkle}{RGB}{130,120,183}
\definecolor{Managed-PineGreen}{RGB}{0,131,108}
\definecolor{Managed-RoyalBlue}{RGB}{0,104,180}

\numberwithin{equation}{section}
\allowdisplaybreaks

\declareslashed{}{/}{0}{0}{\nabla}

\newcommand{\Ibar}{{\widebar I}}
\newcommand{\Jbar}{{\widebar J}}

\usepackage[normalem]{ulem}

\begin{document}

\begin{flushright}
    \today \\
    HU-EP-24/35-RTG
\end{flushright}

\vskip 10mm
\begin{center}
{\LARGE\textbf{ \bf The Dark Side of }}\\
\vskip 5mm
{\LARGE\textbf{ \bf  Double-Tensor Multiplets}}\\
\vskip 1 cm
{L.\ Andrianopoli}$^{[a,b]}$,
{G.\ Casale}$^{[c]}$,
{L.\ Ravera}$^{[a,b,d]}$,
{A. Santambrogio}$^{[e]}$
\end{center}

\vskip 0.3in

\noindent
{\small
$^{[a]}$ {Politecnico di Torino, Corso Duca degli Abruzzi 24, 10129 Torino, Italy} \\
$^{[b]}$ {INFN, Sezione di Torino, Via P. Giuria 1, 10125 Torino, Italy} \\
$^{[c]}$ {Institut f\"ur Physik, Humboldt-Universit\"at zu Berlin, Zum Großen Windkanal 2, \\
\hspace*{0.35cm} 12489 Berlin, Germany} \\
$^{[d]}$ {GIFT -- ``Grupo de Investigaci\'{o}n en Física Te\'{o}rica", \\
\hspace*{0.35cm} Universidad Cat\'{o}lica De La Sant\'{i}sima Concepci\'{o}n, Concepci\'{o}n, Chile} \\ 
$^{[e]}$ {INFN, Sezione di Milano, Via Celoria 16, 20133 Milano, Italy}
}
\vskip 1in
\begin{center}
{\bf{Abstract}}
\end{center}

We explore the properties of a set of free double-tensor multiplets in $\mathcal{N}=2$ supersymmetry, focusing on their behavior within rigid superspace. These multiplets can be obtained {from hypermultiplets} by Hodge-dualizing {half of their} scalars, and 
feature an off-shell matching of bosonic and fermionic degrees of freedom.
Despite this fact, the supersymmetry algebra results to close only on-shell. 
Our analysis is conducted both in superspace, using the geometric (rheonomic) approach, and in spacetime,
comparing how our results are obtained in the two approaches. 
Notably, the  cohomology of superspace requires
that the scalars Hodge-dual to the antisymmetric tensors crucially contribute to  the superspace description of the tensors super-field strengths. This shows an inherent non-locality of the theory, already in the free case, which however does not forbid a  Lagrangian description.

\vfill
\noindent
{\footnotesize{\tt laura.andrianopoli@polito.it};\\
{\tt giuseppe.casale@physik.hu-berlin.de}; \\
{\tt lucrezia.ravera@polito.it}; \\
{\tt alberto.santambrogio@mi.infn.it} }

\numberwithin{equation}{section}
\clearpage

\tableofcontents

\clearpage

\section{Introduction}\label{introsec}

Finding a general \emph{off-shell} formulation of supersymmetric field theories is a major challenge of theoretical physics. It would be a very desirable goal, in particular in the rigid supersymmetric case, towards their description as quantum field theories. 
Still, this is a problematic issue. It is possible  in general for  theories with 4 supercharges ($\mathcal{N}=1$ in four spacetime dimensions): Several  (on-shell equivalent) versions of off-shell representations of $\mathcal{N}=1$ supersymmetry were indeed constructed since the early  years of the birth of supersymmetry \cite{Ferrara:1978em, Wess:1978bu, Stelle:1978ye, Akulov:1976ck, Fradkin:1978jq,Ferrara:1978jt,Sohnius:1981tp, Gates:1981tu,DAuria:1981pyp,DAuria:1982mkx,DAuria:1984glj}. However, in the $\mathcal{N}$-extended supersymmetric cases, an off-shell supersymmetric description \emph{in ordinary superspace} ($\mathcal{M}^{4|4\mathcal{N}}$, in the $D=4$ case) can be formulated only for a very restricted class of supermultiplets \cite{deWit:1979xpv,Fradkin:1979cw,Fradkin:1979as,deWit:1980lyi,Howe:1982tm,deWit:1982na,Galperin:1986fg,Gonzalez-Rey:1997pxs,deWit:2006gn,Lauria:2020rhc}. 
An interesting analysis for the rigid supersymmetric $\mathcal{N}=2$ case of supermultiplets admitting an off-shell extension can be found in \cite{Gates:2014vxa}.

$\mathcal{N}$-extended supersymmetry constrains the dynamics of the fields more strongly with respect to the $\mathcal{N}=1$ case, giving in general better renormalization properties to the corresponding theories. 
Rigid $\mathcal{N}$-extended supersymmetric theories are anyhow studied as quantum field theories on spacetime by expressing their field content either in terms of physical components or in terms of $\mathcal{N}=1$ off-shell superfields \cite{Gates:1983nr}, even though in this way their non-renormalization properties are not fully manifest and the calculations quite involved. This is in general possible because the actions of the supersymmetric theories are off-shell invariant under supersymmetry transformations, even when the  transformations close the superalgebra only on-shell.

Studying $\mathcal{N}$-extended theories in terms of $\mathcal{N}=1$  off-shell  representations allows to partially recover their off-shell structure. However, an important difference between $\mathcal{N}=1$ and  $\mathcal{N}$-extended supersymmetry resides in their different cohomological structure. This  is best caught in their superspace formulation: Indeed, in the $\mathcal{N}$-extended cases the Fierz identities among the gravitini 1-forms (which span the cotangent space of the odd directions of superspace, even in the rigid supersymmetric case) are a source of non-trivial cocycles, thus endowing the extended theories with a much richer cohomological structure with respect to the minimal case. This is responsible, for example, for the  central extension of the supersymmetry algebra, where the central charges are in fact associated with \emph{topological charges} \cite{Witten:1978mh}.
Such richness is however lost in the $\mathcal{N}=1$ formulations of the $\mathcal{N}$-extended 
theories, which then describe, through their quantum Lagrangian,  perturbative patches of the quantum field theory \cite{Seiberg:1994rs,Seiberg:1994aj}.

A general off-shell description of representations for extended supersymmetry, with fully manifest supersymmetry,  was shown to be possible at the cost of including an infinite number of auxiliary fields. 
As discussed in detail in \cite{Galperin:2001seg} for the $\mathcal{N}=2$ and $\mathcal{N}=3$ cases in $D=4$, this was shown to be due to some ``no-go" theorems \cite{Howe:1985ar,Rivelles:1982gn,Bazhanov:1980ku}, stating that an off-shell formulation   of matter multiplets  with a finite number of component fields is not possible when the bosonic physical fields belong to a \emph{non-real} representation of the R-symmetry group. In particular, this is the case for the hypermultiplets in the $\mathcal{N}=2$, $D=4$ theory, which are self-CPT conjugate multiplets with maximum helicity $1/2$ and scalars sitting in the (pseudo-real) fundamental representation of  the  R-symmetry group $SU(2)$.

In the presence of an infinite number of auxiliary fields, instead, the no-go theorems can be evaded and an off-shell formulation found. 
Two main different approaches were developed during the years, which are both formulated  on some extensions of the notion of superspace. They were named, respectively, \emph{harmonic superspace} \cite{Galperin:2001seg} and \emph{projective superspace} \cite{Karlhede:1984vr,Lindstrom:1989ne} approaches, complementary to each other in many respects (see \cite{Kuzenko:2022ajd} for details), yet mathematically related \cite{Kuzenko:1998xm,Jain:2009aj,Butter:2012ta}.

In this paper, trying to better understand the critical issues about this point  towards exploring alternative superspace approaches, we focus on a set of
$\mathcal{N}=2$ multiplets, strictly related to hypermultiplets: the so-called \emph{double-tensor multiplets} \cite{Brandt:2000uw,Theis:2002er,Theis:2003jj,Djeghloul:2012pr,Cribiori:2018xdy}. 
These multiplets have the intriguing peculiarity of exhibiting  the matching of bosonic and fermionic degrees of freedom (d.o.f.) both off- and on-shell.
We consider the simplest case of \emph{free} double-tensor multiplets in rigid superspace, which makes transparent our investigation on their off-shell behavior,   exploring both the cases of flat and super anti-de Sitter (AdS) backgrounds.

The double-tensor multiplets are complex multiplets whose bosonic field content is given by  complex scalars $L^I$ and  complex antisymmetric tensors $B_{\mu\nu}^I$, together with their complex conjugates $\widebar L^\Ibar, \widebar B_{\mu\nu}^\Ibar$ (where the indices $I,\Ibar=1,\ldots ,n$ label the multiplets), while the fermionic field content is composed by Dirac spinors $\chi^I$ (with complex conjugate $\chi^\Ibar \neq \chi^I$). These multiplets can  also be obtained by dualizing half of the scalars of a set of hypermultiplets, in such a way that  $U(n)$ out of the $SU(2)\times Sp(2n)$ Hyper-Kähler holonomy be preserved. They belong to the more general family of \emph{tensor multiplets}, investigated on-shell  at the supergravity level in \cite{Louis:1996ya,DallAgata:2003sjo,Sommovigo:2004vj,DAuria:2004yjt,Sommovigo:2005fk,Andrianopoli:2011zj} and obtained by dualizing an arbitrary number of hyperscalar isometries, but they are special, in that they are the only ones in the family with off-shell matching of degrees of freedom.

In $\mathcal{N}=1$ supersymmetry, analogous multiplets exist, the so-called \emph{linear multiplets} (see, e.g., \cite{Bertolini:1994cb},\cite{Louis:1996ya}). These are real multiplets featuring one real scalar and one real antisymmetric tensor as bosonic field content, and with a Majorana spinor as superpartner. They  can be obtained by Hodge-dualizing the pseudoscalar in an $\mathcal{N}=1$ Wess-Zumino muliplet into an antisymmetric tensor field (a 2-form). The $\mathcal{N}=1$ linear multiplets enjoy off- and on-shell matching of d.o.f. and  close the supersymmetry algebra identically, without the use of the field equations.

We will analyze the $\mathcal{N}=2$ double-tensor multiplets both in superspace, within the geometric (a.k.a. \emph{rheonomic}) approach \cite{Castellani:1991eu} (see also  \cite{DAuria:2020guc,Ravera:2022vjy,Francois:2024rfh,Andrianopoli:2024qwm} for recent comprehensive reviews of this approach), and in the spacetime component approach, showing how the  same  results are obtained in the two formalisms. 
The comparison between the theory developed in superspace and the one restricted to spacetime is useful because it shows that some significant results we obtain, which are better understood in the rheonomic approach -- where they express properties of the cohomology of the  superspace -- are also retrieved as physical conditions in the spacetime component approach. 

In our study, we find a couple of surprises:  First of all,  despite the off-shell matching of bosonic and fermionic degrees of freedom, the double-tensor multiplets are not off-shell multiplets, as the supersymmetry algebra leaving invariant the Lagrangian closes only when the field equations of the spinors are satisfied, i.e. \emph{on-shell}. This fact, which is  an important difference with respect to the $\mathcal{N}=1$ linear multiplet case, is directly observed in both the superspace and the spacetime component approaches. Such result is not pretty new, as it was already noticed in \cite{Brandt:2000uw}. 
But we find another unexpected peculiarity, related to the definition of the super-field strength of the 2-forms in superspace: 
The consistency of the theory, expressed through the cohomology of the superfield  description in superspace, crucially requires to include  into the spectrum of the theory also the scalars whose field strengths are Hodge-dual to the ones of the tensors themselves, namely those scalars of the parent hypermultiplet model that have been dualized to obtain the double-tensor multiplets from a set of hypermultiplets.

This last finding is unexpected and, to our knowledge, new.
Indeed, a necessary condition for  the  hypermultiplet-scalars to be Hodge-dualized into tensors is that they are axionic fields, appearing in the Lagrangian only covered by derivatives \cite{DallAgata:2003sjo,Sommovigo:2004vj,DAuria:2004yjt,Sommovigo:2005fk,Andrianopoli:2011zj}. As such, they  are expected to  correspond to translational  symmetries 
of the hyper-scalar sector. The same is also implicitly assumed to hold, \emph{a fortiori}, in the theory after dualization, that is for the double-tensor multiplets.
However, we find that those scalars contribute necessarily, not covered by derivative, to the definitions of the tensor super-field strengths in superspace, and they also appear in the superspace Lagrangian. Still, they do not spoil the possibility of a Lagrangian description of the theory, since their superspace field equations are identically satisfied, with no further constraints, upon use of the on-shell conditions on all the physical fields.

The need to include  the free scalars Hodge-dual to the tensor fields 
suggests a possible interpretation to the infinite number of auxiliary fields required  for an off-shell description, and which is a crucial ingredient of the harmonic superspace formulation of these multiplets: They are possibly related to the harmonic expansion of the scalars that have been dualized to obtain the double-tensor multiplets, on-shell non-locally related  to their Hodge-dual tensors.

These peculiar features, related to an inherent non-locality of the theory,  are caught in the simplest way from the superspace description of the theory within the geometric approach, due to the fact  that the Fierz identities on polynomials of the gravitini 1-forms are  sources of non-trivial cocycles which enrich the cohomological structure of superspace with respect to that of spacetime.  
This was noticed in the pioneering work \cite{DAuria:1982uck}, where this feature was applied to construct $D=11$ supergravity in terms of a so-called \emph{Free Differential Algebra}, and it was then further investigated more recently (see, for example, \cite{Andrianopoli:2016osu,Andrianopoli:2017itj,Cremonini:2022cdm,Andrianopoli:2024qwm,Ravera:2018vra,Giotopoulos:2024ovz}). The non-trivial cocycles due to  the odd sector of superspace are recovered also on the restriction to spacetime of the given supergravity theory. However, in the limit to rigid supersymmetry, at least for the case of flat, Minkowskian superspace, the spacetime projection of the gravitino 1-form vanishes, $\Psi_{\mu A}=0$, so that this cohomological richness is not manifest. Still, as we will show explicitly, the non-trivial cocycles from the odd sector of superspace also affect the spacetime projection. Their effects are captured by the emergence also in the restriction to spacetime, for a consistent formulation of the theory, of the scalars dual to the antisymmetric tensors of the multiplet  together with the tensors themselves.

A known effect of the  non-trivial cocycles in superspace is that  they manifest themselves as central (or quasi-central, in higher dimensions) charges  providing, in the $\mathcal{N}$-extended theories, a central extension of the supersymmetry algebra \cite{Witten:1978mh}.
To explore the role of central charges in this context, we then also  analyse if it is possible to embed  our set of double-tensor multiplets in a curved (AdS) rigid supersymmetric background.
Rigid supersymmetric models in curved backgrounds \cite{Festuccia:2011ws,Dumitrescu:2012ha,Dumitrescu:2012at} were not very much explored in the literature, in particular in the case of extended theories (some relevant results in the extended supersymmetric case are \cite{Festuccia:2018rew,Festuccia:2019akm,Festuccia:2020xtv},
and \cite{Andrianopoli:2021sdx} for an application in $D=3$ with $8$ supercharges). Here we find that for the $\mathcal{N}=2$ hypermultiplets it is doable, and we present its explicit superspace construction once  the deformations due to the cosmological constant are included. On the other hand, in the case of the double-tensor multiplets, and in the absence of gauge matter multiplets, 
it appears not to be possible to 
find a consistent description of them in an AdS background, despite the fact that the construction is possible at the supergravity level \cite{Bagger:1983tt,DAuria:1990qxt,Andrianopoli:1996vr,Andrianopoli:1996cm,DallAgata:2003sjo,Sommovigo:2004vj,DAuria:2004yjt,Sommovigo:2005fk,Andrianopoli:2011zj,Trigiante:2016mnt}.

The paper is organized as follows: In Section \ref{fieldcontsect} we give the field content of the flat $\mathcal{N}=2$ background and of the double-tensor multiplets, and we show  how the latter can be obtained from the hypermultiplets via dualization. In Section \ref{superspacemodel} we explicitly construct and analyze the double-tensor model in superspace, adopting the rheonomic approach. We provide the supersymmetry transformation rules, the superspace Lagrangian and the field equations of the theory, showing that: (i) The supersymmetry transformations leaving the {superspace} Lagrangian invariant {only close on-shell}; (ii) It is necessary to include the (free) scalars whose field strengths are Hodge-dual to the ones of the tensors themselves (and complex conjugates) for the consistency of the theory. In Section \ref{spacetimemodel} we give the spacetime component description of the model, showing how these results {are recovered in this framework}. Subsequently, in Section \ref{cosmconstadd} we discuss the case of a curved (super {AdS}) background with negative cosmological constant, presenting the deformation induced by the background for the hypermultiplets and proving that such deformation is not compatible with the {free} double-tensor multiplets. Section \ref{concl} is devoted to our conclusions and possible future developments. In Appendix \ref{conventions} we collect our notation and conventions, together with useful formulas on the Clifford algebra and Fierz identities.   In Appendix 
 \ref{superalgebraapp} we give the $\mathcal{N}=2$ super-Poincar\'e Lie algebra in its standard formulation and in its dual Maurer-Cartan description in terms of 1-forms.  In Appendix \ref{linN1comparison} we make some comparison with the $\mathcal{N}=1$ linear multiplets, which close off-shell.

\section{Field content}\label{fieldcontsect}
In this section we are going to present the field content of the  $\mathcal{N}=2$  double-tensor multiplets in  four-dimensional spacetime, which are the main object of this paper. We will work in rigid $\mathcal{N}=2$ superspace, with R-symmetry $SU(2)\times U(1)$, and we will then project the results on spacetime.

\subsection{The flat $\mathcal{N}=2$ background}

The $\mathcal{N}=2$ background superspace is spanned by the supervielbein 1-form 
\begin{align}
    (V^a,\Psi_A)\,,
\end{align}
where the 1-form $V^a$, with $a=0,1,2,3$, spans the even directions of superspace,  and $\Psi_A$, with $A=1,2$, are instead a couple of Majorana spinor 1-forms spanning the odd directions. 
We will find it convenient to define their Weyl projections with respect to left/right projectors   $\mathbb{P}_\pm \equiv \frac12 (\mathbb{I}\pm\gamma_5)$ as
\begin{align}
    \psi_A \equiv \mathbb{P}_+\Psi_A\,,\qquad
 \psi^A  \equiv \mathbb{P}_-\Psi_A\,,\label{chirpsi} 
\end{align}    
such that  $\gamma_5\psi_A=\psi_A$, $\gamma_5\psi^A=-\psi^A$.  
A complete set of definitions is given in Appendix \ref{conventions}.

The $\mathcal{N}=2$ background also includes a graviphoton 1-form which, however, does not play any role here, since in the flat background the matter multiplets are not charged under it. {Its role is instead relevant in the case of a curved supersymmetric background, as we are going to see in Section \ref{cosmconstadd}.} 

\subsection{The double-tensor multiplets}

The double-tensor multiplets are given by $n$  complex multiplets, labeled by  indices $I,\Ibar=1,\ldots ,n$:
\begin{align}
    \upphi^I\equiv\left(L^I, B^I, \chi^I
\right)\,,\quad \widebar\upphi^\Ibar\equiv\left(\widebar L^\Ibar, \widebar B^\Ibar, \chi^\Ibar
\right)\,,
\end{align}
where $L^I$  are $n$ holomorphic scalars, $B^I = B^I_{\mu\nu}dx^\mu\wedge dx^\nu$ are $n$ holomorphic 2-forms, while $\chi^I$ are  $n$ Dirac spinors, defined by
\begin{equation}
    \chi^I = \zeta^I + i \zeta_\Ibar \delta^{I\Ibar}\,.\label{chii}
\end{equation}
Here, \begin{equation}\zeta_\upalpha = \gamma_5 \zeta_\upalpha \text{ and } \zeta^\upalpha = -\gamma_5 \zeta^\upalpha \,, \label{hyperinos}
\end{equation}
where $\upalpha=(I,\Ibar)$, appear in the left and right projections of the Dirac spinors:
\begin{equation}\mathbb{P}_- \chi^I = \zeta^I\,,\quad \mathbb{P}_+ \chi^I = i\zeta_\Ibar \delta^{I\Ibar}\,.\end{equation}

The complex conjugate multiplet $\widebar\upphi^\Ibar$
is given in terms of the  {bosonic} fields $\widebar L^\Ibar= (L^I)^*$, 
$\widebar B^\Ibar  = (B^I)^*$, which are $n$ anti-holomorphic scalars and 2-forms, respectively, and by the  spinors $\chi^\Ibar${:}\footnote{In general, we denote with a ``bar" on  spinors the adjoint spinors, as defined in Appendix \ref{conventions}. For this reason, we omit the bar on the Dirac spinors $\chi^\Ibar$ belonging to the conjugate multiplets $\widebar\upphi^\Ibar$.} 
\begin{equation}
    \chi^\Ibar = \zeta_I\delta^{I\Ibar} + i \zeta^\Ibar \,.\label{chiibar}
\end{equation}

The adjoint spinors are:
\begin{subequations}
\begin{align}
\widebar\chi^\Ibar \equiv (\chi^I)^\dagger \gamma^0 &=   \bar\zeta_I \delta^{I\Ibar}- i \bar\zeta^\Ibar \,, \\
  \widebar\chi^I \equiv (\chi^\Ibar)^\dagger \gamma^0 &=      \bar\zeta^I - i \bar\zeta_\Ibar \delta^{I\Ibar} \,,
\end{align}
\end{subequations}
where $\bar\zeta_\upalpha =+ \bar\zeta_\upalpha  \gamma_5$ and $\bar\zeta^\upalpha = -\bar\zeta^\upalpha  \gamma_5$ give the chiral projections of the adjoint spinors, so that:
\begin{subequations}\begin{align}
    \widebar\chi^\Ibar \mathbb{P}_+ = \bar\zeta_\Ibar \delta^{I\Ibar} \,&, \quad 
    \widebar\chi^\Ibar \mathbb{P}_- = -i\bar\zeta^\Ibar \,, \\
    \widebar\chi^I \mathbb{P}_+ = - i \bar\zeta_\Ibar \delta^{I\Ibar} \,&, \quad 
    \widebar\chi^I \mathbb{P}_- = \bar\zeta^I\,.
\end{align}\end{subequations}
Note that the positions of the indices are immaterial in the bosonic sector, so that indices of a given holomorphy ($I$, or $\Ibar$) can be freely raised and lowered with the Kronecker delta $\delta_{IJ}$, or $\delta_{\Ibar\Jbar}$, while for the {chiral projections of the} spinors they are stuck, being associated with the chirality properties of the given component.

\subsubsection*{Counting the degrees of freedom}

The total number of bosonic degrees of freedom in each double-tensor multiplet is  given as follows:  The tensor field $B^I_{\mu\nu}$ has 3 off-shell complex d.o.f., that on-shell reduce to 1 complex d.o.f.; The scalar $L^I$ has 1 complex d.o.f. both off- and on-shell. The total number of bosonic d.o.f is then given by  8 real  d.o.f. off-shell and 
4 real  d.o.f.  on-shell.\\
On the other hand, the fermionic field content of each multiplet  being given by Dirac spinors, the total number of fermionic off-shell d.o.f. amounts to 4 complex, namely 8 real, d.o.f. that  are halved on-shell, so that we have on-shell 4 real degrees of freedom.\\
The double-tensor multiplets thus enjoy exact matching of bosonic and fermionic d.o.f., both off-shell and on-shell. In the $\mathcal{N}=1$ case, multiplets with similar features exist, the so-called linear multiplets \cite{Bertolini:1994cb},\cite{Louis:1996ya}, each featuring a real scalar and a real antisymmetric tensor  (3+1 off-shell real boson d.o.f.), together with one Majorana spinor (4 off-shell real fermionic d.o.f.). The linear multiplets  enjoy matching of d.o.f. both off- and on-shell, and they do close the supersymmetry algebra without use of the field equations. 
As we are going to show in Section \ref{superspacemodel}{, for the $\mathcal{N}=2$ double-tensor multiplets, instead, despite } their ``power-counting" matching, closure of the supersymmetry algebra  requires use of the spinor field equations, and is therefore satisfied only on-shell. This fact was already observed in \cite{Brandt:2000uw}.

The difference between the $\mathcal{N}=1 $ and the $\mathcal{N}=2 $ case resides in the different cohomological structure of the two superspaces,  expressed through their different Fierz identities. A detailed comparison of off-shell behavior of the double-tensor multiplets with the $\mathcal{N}=1 $ linear multiplets is
given in Appendix \ref{linN1comparison}.

\subsection{Dualizing hypermultiplets}\label{hypers}

Double-tensor multiplets can be obtained by Hodge-dualizing half of the scalars in a set of $n$ hypermultiplets, in such a way to preserve one of the three complex structures of their Hyper-K\"{a}hler structure, and thus breaking  the manifest $SU(2)$ R-symmetry: 
\begin{align}
    SU(2)\to U(1)\,.
\end{align}
 {The h}ypermultiplets are $D=4$, $\mathcal{N}=2$ self-CPT conjugate multiplets, given in terms of the fields
\begin{align}
    (q^{u}, \zeta_\upalpha, \zeta^\upalpha)\,,
\end{align}
where $u=1,\ldots ,4n$   and $\upalpha=1,\ldots,2n$ is an $Sp(2n,\mathbb{R})$ index. 
A set of $n$ hypermultiplets  is therefore composed by $4n$ real scalars,  $q^{u}$, which can be thought of as coordinates of an Hyper-K\"ahler $\sigma$-model,
  and by $2n$ Majorana spinors, which can be decomposed in their chiral left-handed and right-handed projections defined by eq. \eqref{hyperinos}.\footnote{{Let us stress that   $\upalpha$  is not a  spinorial index, which is  left implicit here, but is instead an index related to the internal symmetry  $Sp(2n)$, which enumerates the fields in the hypermultiplets. In our notation, the  position  {of the internal symmetry index } on the spinor (lower or upper) is also associated with its chirality, as explained in Appendix \ref{conventions}.}} 
The metric $h_{uv}$ on the Hyper-K\"ahler $\sigma$-model can be conveniently expressed as 
$h_{uv}= \mathcal{U}^{A\upalpha}_u\mathcal{U}^{B\upbeta}_v\epsilon_{AB}\mathbb{C}_{\upalpha\upbeta}$,  in terms of a complex scalar vielbein $\mathcal{U}^{A\upalpha}(q)= \mathcal{U}^{A\upalpha}_u dq^u$, with   holonomy in $SU(2)\times Sp(2n,\mathbb{R})$, and enjoying the reality condition
\begin{equation}
\mathcal{U}_{A\upalpha}=\left(\mathcal{U}^{A\upalpha}\right)^*= \epsilon_{AB}\mathbb{C}_{\upalpha\upbeta}
\mathcal{U}^{B\upbeta}\,. \label{realu}
\end{equation}
Here, $A=1,2$ is an $SU(2)$ R-symmetry index, $\epsilon_{AB}= -\epsilon_{BA}$, and $\mathbb{C}_{\upalpha\upbeta}= -\mathbb{C}_{\upbeta\upalpha}$ is the invariant metric of $Sp(2n,\mathbb{R})$.

We want to analyze the conditions for  the dualization of half of the scalars in the hypermultiplets, in such a way to preserve a complex structure in the residual scalar sector. This requires, as a preliminary condition, the Hyper-K\"{a}hler manifold spanned by the scalars $q^u$ to admit  at least $2n$ translational isometries.

To find the correct assignment in order to get  complex multiplets as the resulting multiplets, it is useful to explicitly  decompose the symplectic indices as $\upalpha=(I,\widebar I)$, $I=1,\ldots n$, $\widebar I=1,\ldots n$, and write 
\begin{align}
    \mathbb{C}_{\upalpha\upbeta}= \begin{pmatrix}
\mathbb{O}& \delta_{I\Ibar }\cr
-\delta_{I\Ibar} &\mathbb{O}
\end{pmatrix}\,,\quad \epsilon_{AB}= \begin{pmatrix}
0&1\cr-1&0
\end{pmatrix}\,,
\end{align}
so that  the reality condition \eqref{realu}
can be rewritten as
\begin{align}
\mathcal{U}_{1I}=\left(\mathcal{U}^{1I}\right)^*= 
\mathcal{U}^{2\Ibar}\delta_{I\Ibar}\,,\quad \mathcal{U}_{2I}=\left(\mathcal{U}^{2I}\right)^*= 
-\mathcal{U}^{1\Ibar}\delta_{I\Ibar}\,.
\end{align}

We consider the case where the metric on the scalar $\sigma$-model is flat, that is $\mathcal{U}^{A\upalpha}_u=\delta^{A\upalpha}_u$.\footnote{This is always possible, in general, in the rigid supersymmetric case, with a proper choice of coordinates on the $\sigma$-model.} With this choice, corresponding to choose as scalar coordinates the complex fields $(L^I,M^I)$ and their complex conjugates $(\widebar L^\Ibar,\widebar M^\Ibar)$ such that
\begin{align}
    q^{A\upalpha}: \,\quad \left\{q^{1I}=L^I\,,\quad q^{2I}=i\,M^I\,,\quad 
q^{1\Ibar}=i\,\widebar M^\Ibar\,,\quad q^{2\Ibar}=\widebar L^\Ibar\right\}\,,
\end{align}
we have
\begin{equation}   \label{flatsm}   \mathcal{U}^{1I}= dL^I= (\mathcal{U}_{2\Ibar})^*\,,\quad \mathcal{U}^{2I}= i\,dM^I = -(\mathcal{U}_{1\Ibar})^*\,.
\end{equation}
To obtain the double-tensor multiplets, the scalar  vielbein $\mathcal{U}^{2 I}$, together with its complex conjugate, $-\mathcal{U}^{1 \Ibar}$,  are Hodge-dualized into   3-form field strengths $H^I, \widebar H^\Ibar$ of  {the} complex 2-forms $B^I\equiv B^I_{\mu\nu}dx^\mu\wedge dx^\nu$, $\widebar B^\Ibar \equiv \widebar B^\Ibar_{\mu\nu}dx^\mu\wedge dx^\nu$:
\begin{equation}
    \star\,\mathcal{U}^{2 I}= {\frac 32}\,H^I \,,\quad 
\star\,\mathcal{U}^{1 \Ibar}= -{\frac 32}\,\widebar H^\Ibar \,,
\end{equation}
that is
\begin{subequations}
\label{dual}
\begin{align}
{\frac 23}\epsilon^{\mu\nu\rho\sigma}\, \mathcal{U}^{2 I}_u\partial_\sigma q^u&= H^{I | \mu\nu\rho}=  \partial^{[\mu}B^{I | \nu\rho]}\,,\\
-{\frac 23}\epsilon^{\mu\nu\rho\sigma}\, \mathcal{U}^{1 \Ibar}_u\partial_\sigma q^u&= \widebar H^{\Ibar | \mu\nu\rho}=  \partial^{[\mu}\widebar B^{\Ibar | \nu\rho]}\,.
\end{align}
\end{subequations}
We remark that the dualization \eqref{dual} breaks the manifest $SU(2)\times Sp(2n)$ covariance of the $\mathcal{N}=2$ model considered.

The {remaining} scalar sector  is spanned by the complex vielbein $E^I\equiv \mathcal{U}^{1 I} = dL^I $   and by its complex conjugate $\widebar E^\Ibar\equiv \mathcal{U}^{1 \Ibar} = d\bar L^\Ibar$. 

To identify which components of the hyperini belong to the multiplet $\upphi^I$ and which ones to its complex conjugate $\widebar\upphi^\Ibar$,
we can consider the supersymmetry transformation laws of the scalars $L^I, \widebar L^\Ibar$ in the hypermultiplets prior to dualizations (for example, see \cite{Andrianopoli:1996vr}).
From the general relation 
\begin{equation}
\mathcal{U}^{A\upalpha}_u\delta_\varepsilon q^u = \widebar \varepsilon^A\zeta^\upalpha +\epsilon^{AB}\mathbb{C}^{\upalpha\upbeta}\widebar\varepsilon_B\zeta_\upbeta\,,   
\end{equation}
where $\varepsilon_A=\mathbb{P}_+\varepsilon_A$, $\varepsilon^A=\mathbb{P}_-\varepsilon^A$ are respectively the  left and right projections of the supersymmetry parameter, we find:
\begin{subequations}\begin{equation}\begin{split}
 \delta_\varepsilon L^I=&\delta_\varepsilon q^{1I} = \widebar \varepsilon^1\zeta^I+\widebar\varepsilon_2\zeta_\Ibar \delta^{I\Ibar} \\
 =&\left(\widebar\varepsilon^1 -i\,\widebar\varepsilon_2\right)\chi^I\,,\\
\end{split}\end{equation}\begin{equation}\begin{split}
 \delta_\varepsilon \widebar L^\Ibar=&\delta_\varepsilon q^{2\Ibar} = \widebar \varepsilon_1\zeta_I\delta^{I\Ibar}+\widebar\varepsilon^2\zeta^\Ibar \\
 =&\left(\widebar\varepsilon_1 -i\,\widebar\varepsilon^2\right)\chi^\Ibar\,,
\end{split}\end{equation}\end{subequations}
where $\chi^I, \chi^\Ibar$ are the Dirac spinors introduced in \eqref{chii}, \eqref{chiibar}.

\section{Construction of the double-tensor multiplets in superspace}\label{superspacemodel}

In {this}  section we will present our results in the rheonomic approach in superspace, where they have been first derived and where they are fully manifest.

\subsection{Rheonomic approach in superspace}

We  follow the general prescriptions of the rheonomic approach \cite{Castellani:1991eu} (see also the recent reviews \cite{DAuria:2020guc,Andrianopoli:2024qwm}), where the supersymmetry transformation laws of the fields in the multiplet are obtained as diffeomorphisms along the odd directions of superspace. 

The procedure {goes} as follows: One first defines the super-field strengths in superspace of all the fields involved (both dynamical and background fields) in terms of differential forms and their exterior differential operator $d$, satisfying $d^2=0$, based on symmetry arguments only --
for the supergravity background, the super-field strengths are vanishing and defined by the Maurer-Cartan equations of the $\mathcal{N}=2$ super-Poincaré algebra in its dual form; for the dynamical fields of the double-tensor multiplets, they are instead non-vanishing, and determined by Lorentz-covariant expressions in superspace required to admit, as a possible configuration, the vacuum configuration where all the field strengths are zero.
{The $\mathcal{N}=2$ supersymmetry algera and its dual description in terms of Maurer-Cartan forms is given in Appendix \ref{superalgebraapp}.}

By acting with the exterior differential on the super-field strengths, we  obtain  the Bianchi identities that the fields have to satisfy. One then requires   the field strengths to be differential forms \emph{in superspace}, by giving them a general parametrization  on a basis of the cotangent space to superspace, spanned by the supervielbein $(V^a,\Psi_A)$.  The actual expressions of the various components in the parametrizations are then determined by requiring that  the parametrizations of the super-field strengths do satisfy the Bianchi identites in superspace. 

\medskip

The flat $\mathcal{N}=2$ superspace  background is given in terms of the vanishing Lorentz curvature $R^{ab}$,  supertorsion $T^a$ and  gravitini field strengths $\rho_A$, $\rho^A$. The background also includes   the (vanishing) graviphoton super-field strength, $F^0$. They satisfy the Maurer-Cartan equations of the super-Poincaré algebra:\footnote{Here, $\mathcal{D}$ denotes the Lorentz-covariant derivative, acting on Lorentz vectors $K^a$ as $\mathcal{D}K^a\equiv dK^a+\omega^{a}{}_b \wedge K^b$, and on spinors $\lambda$ as $\mathcal{D}\lambda\equiv d\lambda +\frac 14 \gamma_{ab}\,\omega^{ab}\wedge\lambda$. On the flat background, where $\omega^{ab}$ is a pure gauge, they are gauge-equivalent to the ordinary exterior differentials. To lighten the notation, we will generally omit writing the wedge product ``$\wedge$" between differential forms.}
\begin{subequations}\begin{align}
    \mathcal{R}^{ab} &\equiv d\omega^{ab} + \omega^a{}_c\wedge\omega^{cb} = 0 \,, \\
    T^a&\equiv \mathcal{D}V^a  -i \widebar\psi_A \gamma^a \psi^A=0 \,, \\
    F &\equiv dA  - \frac 1{\sqrt 2} \widebar\psi_A\psi_B\epsilon^{AB} - \frac 1{\sqrt 2} \widebar\psi^A\psi^B\epsilon_{AB}=0  \,, \\
    \rho_A &\equiv \mathcal{D}\psi_A =0 \,, \\
    \rho^A &\equiv \mathcal{D}\psi^A =0 \,.
\end{align}\end{subequations}
The super-field strengths of the double-tensor multiplets, in the absence of other matter multiplets, are defined as  Lorentz-covariant expressions admitting the vacuum configuration in superspace, and they are 
\begin{subequations}
\label{def}
\begin{align}
    E^I &\equiv dL^I\,, \\
    H^I
    &\equiv dB^I +\left[8L^I\widebar\psi_1\gamma_a\psi^2  -4\,i\, M^I \left(\widebar\psi_1\gamma_a\psi^1-\widebar\psi_2\gamma_a\psi^2\right)\right]V^a \,, \label{defH} \\
\mathcal{D}\chi^I&\equiv  d\chi^I+ \frac 14 \gamma_{ab}\omega^{ab}\chi^I\,,
\end{align}
\end{subequations}
together with the ones of their complex conjugates,
\begin{subequations}
\label{defb}
\begin{align}
\widebar E^\Ibar &\equiv d\widebar L^\Ibar \,,
\\
\widebar H^\Ibar &\equiv  d\widebar B^\Ibar -\left[8\widebar L^\Ibar\widebar\psi^1\gamma_a\psi_2 +4\,i\, \widebar M^\Ibar \left(\widebar\psi_1\gamma_a\psi^1 
-\widebar\psi_2\gamma_a\psi^2 \right)\right]V^a \,, \label{defHb}\\
\mathcal{D}\chi^\Ibar&\equiv  d\chi^\Ibar+ \frac 14 \gamma_{ab}\omega^{ab}\chi^\Ibar\,.
\end{align}
\end{subequations}

We note that the definitions \eqref{defH} and \eqref{defHb} include new objects that we name ${M^I}$. 
The corresponding 2-gravitini monomials also appear in the (on-shell) formulations of  hypermultiplets, at the supergravity level, where an arbitrary number of scalars are dualized into 2-form fields \cite{Andrianopoli:1996vr,Andrianopoli:1996cm,DallAgata:2003sjo,Sommovigo:2004vj,DAuria:2004yjt,Sommovigo:2005fk,Andrianopoli:2011zj}. In these papers, such terms multiply the components of the $SU(2)$-composite connection on the scalar $\sigma$-model, and are assumed to be functions only  of the scalars remaining after dualization (in our notations, the $L^I$), but not of the scalars that have been dualized. As we are going to see, this last statement has to be reconsidered, though. 

The super-field strengths defined above should satisfy the following Bianchi identities in superspace:
\begin{subequations}\label{bianchiid}\begin{align}
    dE^I &= 0  \,,\\
    dH^I&= \left[8dL^I\widebar\psi_1\gamma_a\psi^2  -4\,i\, dM^I \left(\widebar\psi_1\gamma_a\psi^1-\widebar\psi_2\gamma_a\psi^2\right)\right]V^a \,,
    \label{dh}
    \\
     \mathcal{D}^2 \chi^I&=  0\,, \quad \text{that is } \begin{cases}
        \mathcal{D}^2 \zeta^I&=0\cr
        \mathcal{D}^2 \zeta_{\widebar I}&=0\cr   
    \end{cases} \,,
   \end{align}\end{subequations}
and
\begin{subequations}\label{bibar}\begin{align}
    d \widebar E^\Ibar &= 0  \,,\\
    d\widebar H^\Ibar&= - \left[ 8d\widebar L^\Ibar\widebar\psi^1\gamma_a\psi_2 +4\,i\, d\widebar M^\Ibar \left(\widebar\psi_1\gamma_a\psi^1 -\widebar\psi_2\gamma_a\psi^2 \right)\right]V^a \,, \label{dhb} \\  
    \mathcal{D}^2 \chi^\Ibar&= 0\,, \quad \text{that is } \begin{cases}
        \mathcal{D}^2 \zeta_I&=0\cr
        \mathcal{D}^2 \zeta^{\widebar I}&=0\cr   
    \end{cases}\,.
\end{align}\end{subequations}
Consistency of the theory in superspace {relies on Fierz identities in the odd directions of superspace \footnote{The relevant Fierz identities are given in Appendix \ref{Fierzapp}.} and} requires that the  Bianchi identities \eqref{bianchiid}, \eqref{bibar} have to be satisfied when the super-field strengths are expressed as exterior forms, on a basis of superspace spanned by the (flat) supervielbein $(V^a,\Psi_A)$.

We find that 
this requires the supercurvatures to enjoy the following parametrizations on a basis of 1- and 2-forms in superspace:
\begin{subequations}
\label{param} 
\begin{align}
    E^I &= E^I_a V^a +\widebar\psi^1\zeta^{I}+\widebar\psi_2 \zeta_{\widebar I}\delta^{I\widebar I}  \,, \label{epar}\\  
    H^I &= H^I_{ abc}V^aV^bV^c-2 i  \left(\widebar\psi_1   \gamma_{ab}\zeta_{\widebar J}\delta^{I\widebar J}+\widebar\psi^2\gamma_{ab}\zeta^{I} \right)V^aV^b \,, \label{hpar}\\
    \mathcal{D}\zeta_{\Ibar} &= \mathcal{D}_a \zeta_{\Ibar} V^a  +\delta_{I\Ibar}\left(i E^I_a \gamma^a\psi^2 -\frac 32 \,i h^I_a \gamma^a\psi^1\right) \,, \\
    \mathcal{D}\zeta^I &= \mathcal{D}_a \zeta^I V^a + i E^I_a \gamma^a\psi_1 + \frac 32\,i h^I_a \gamma^a\psi_2 \,,
    \label{SupSpParametrMajorana}
\end{align}
\end{subequations}
and
\begin{subequations}
\label{paramb} 
 \begin{align}
   \widebar E^\Ibar &= \widebar E^\Ibar_a V^a +\widebar\psi_1\zeta_{I}\delta^{I\widebar I}+\widebar\psi^2 \zeta^{\widebar I} \,,\\  
    \widebar H^\Ibar &= \widebar H^\Ibar_{ abc}V^aV^bV^c + 2 i \left(\widebar\psi^1   \gamma_{ab}\zeta^{\widebar I}+\widebar\psi_2\gamma_{ab}\zeta_{J} \delta^{J\widebar I}\right)V^aV^b \,, \\
    \mathcal{D}\zeta^{\widebar I} &= \mathcal{D}_a \zeta^{\widebar I} V^a  +i \widebar E^\Ibar_a \gamma^a\psi_2 -  \frac 32\,i\, \bar h^\Ibar_a \gamma^a\psi_1 \,, \\
    \mathcal{D}\zeta_I &= \mathcal{D}_a \zeta_I V^a +\delta_{I\widebar I}\left( i \widebar E^\Ibar_a \gamma^a\psi^1 + \frac 32\,i\, \bar h^\Ibar_a \gamma^a\psi^2  \right) \,.
    \label{SupSpParametrMajoranacc}
\end{align}  
\end{subequations}
In order for the parametrizations \eqref{param} of the supercurvatures to satisfy the Bianchi identities \eqref{bianchiid}, the tensors appearing in the parametrizations of the spinors should be identified as follows in terms of the physical fields:
\begin{align}
   E^I_{a} &= \partial_a L^I \,, \label{relEL} \\
    \widebar{E}^\Ibar_{a} &= \partial_a \widebar L^ {\Ibar } \,, \label{relELbar}
\end{align}
and
\begin{align}
    h^I_{a} &= \frac 16 \epsilon_{abcd}H^{I|bcd} = {\frac16} \epsilon_{abcd} \partial^b B^{I|cd}  \,, \label{relhH} \\
    \widebar{h}^\Ibar_{a} &= \frac 16 \epsilon_{abcd}\widebar{H}^{\Ibar|bcd} = {\frac16} 
 \epsilon_{abcd} \partial^b \widebar B^{\Ibar|cd} \,. \label{relhHbar}
\end{align}
From the study of the $d^2$-closure of the Bianchi identities, we also find the following relations:
\begin{subequations}\begin{align}
    &\partial_{[a}E^I_{b]} = 0 \,, \\
    &{\buildrel \psi  \over \nabla } 
 E^I_a=\widebar\psi^1\partial_a\zeta^I + \widebar\psi_2\partial_a\zeta_{\widebar I}\delta^{I\widebar I}\,,
    \\
    & \partial_{[d}H_{I\,abc]} = 0\,, \\
    &{\buildrel \psi  \over \nabla }  H^I_{abc}=-2 i 
    \widebar\psi^2 \gamma_{[ab}\partial_{c]}\zeta^I -2 i 
    \widebar\psi_1 \gamma_{[ab}\partial_{c]}\zeta_{\widebar J} \delta^{I\widebar J}\,,
\end{align}\end{subequations}
where we have denoted with ${\buildrel \psi  \over \nabla } \equiv \widebar\psi^A\nabla_A + \widebar\psi_A\nabla^A$ the spinorial derivative, that is the components of the differential along the odd directions of superspace.

Moreover, we find that the Bianchi identities are satisfied only if we require:
\begin{equation}
    \slashed\partial {\zeta}^I=0\,,\quad \slashed\partial {\zeta}_\Ibar=0\,,
\end{equation}
that is when the (free) spinor-field equations are satisfied.
As a consequence, the double-tensor multiplets are well-defined representations of supersymmetry in superspace only on-shell, namely they close the supersymmetry algebra only on-shell. 

This is not enough, however. Indeed, we find that the scalar quantities $M^I$, $\widebar M^\Ibar$ are not functions of the scalars $L^I$, $\widebar L^\Ibar$ of the supermultiplets, but should be new scalars satisfying:
\begin{subequations}\label{dm}
\begin{align}   dM^I &= \partial_a M^I V^a+i \left(\widebar\psi_1 \zeta_\Ibar \delta^{I\Ibar} -  \widebar\psi^2\zeta^I\right)\,,
\\    d\widebar M^\Ibar &= \partial_a \widebar M^\Ibar V^a-i \left( \widebar\psi^1\zeta^\Ibar-\widebar\psi_2\zeta_I \delta^{I\Ibar}\right) \,.
\end{align}
\end{subequations}
As we are going to see in the next paragraph, the above {expressions} are nothing but the superspace parametrizations of the scalars that have been Hodge-dualized into the  tensors $B^I_{\mu\nu}, \widebar B^\Ibar_{\mu\nu}$, to get the double-tensor multiplets from a set of hypermultiplets (see eqs. \eqref{hypsca}).

\subsubsection*{Comparison with hypermultiplets in $\mathcal{N}=2$ superspace}

On-shell, the hyperscalars in $\mathcal{N}=2$ superspace 
can be parametrized on the basis of supervielbein  1-forms $(V^a,\psi_A,\psi^A)$  as
\begin{equation}\begin{split}
    \mathcal{U}^{A\upalpha} 
    &= \mathcal{U}^{A\upalpha}_aV^a+ \widebar\psi^A \zeta^\upalpha + \epsilon^{AB}\mathbb{C}^{\upalpha\upbeta}\widebar\psi_B \zeta_\upbeta \,.
\end{split}\end{equation}
Decomposing the $SU(2)$ and symplectic indices we get
\begin{subequations}\begin{align}
    \mathcal{U}^{1 I}&= \mathcal{U}^{1 I}_aV^a+  \widebar\psi^1\zeta^I +  \widebar\psi_2 \zeta_\Ibar \delta^{I\Ibar} =(\mathcal{U}^{2 \widebar I})^*\,\delta^{I\Ibar} \,, \label{u1}\\
    \mathcal{U}^{2 I}&= \mathcal{U}^{2 I}_aV^a+  \widebar\psi^2\zeta^I-  \widebar\psi_1 \zeta_\Ibar \delta^{I\Ibar} =-(\mathcal{U}^{1\Ibar})^* \,\delta^{I\Ibar} \,. \label{u2}
\end{align}\end{subequations}
The above expressions are useful because they show which combinations of the hyperini are the supersymmetric partners of a given component of the scalar vielbein.

As we have discussed above, the double-tensor multiplets  considered here 
exhibit, as bosonic field strengths, a scalar sector spanned by the complex vielbein $E^I\equiv \mathcal{U}^{1 I} $ and a tensorial sector  
obtained after dualizing the scalar vielbein $\mathcal{U}^{2 I}$  into the complex 3-form field strengths $H^I$ of the 2-forms $B^I$, in the parent hypermultiplets model. 
The complex conjugate multiplet features, instead, as bosonic field strengths, the scalar vielbein $\widebar E^\Ibar$ and the 3-forms $\widebar H^\Ibar$, related by Hodge-duality to the scalar vielbein $\widebar{\mathcal{U}}^{1\Ibar}$. 
Altogether,  the superspace parametrizations of the hypermultiplet  scalar vielbein, with   split indices $SU(2)\times Sp(2n) \to U(n)$, read:
\begin{subequations}\label{hypsca}
\begin{align}
    E^I\equiv\;\mathcal{U}^{1 I} &= \mathcal{U}^{1 I}_aV^a+  \widebar\psi^1\zeta^I +  \widebar\psi_2 \zeta_\Ibar \delta^{I\Ibar} \,, \\\label{imi}
    i\,dM^I\equiv\;  \mathcal{U}^{2 I} &= \mathcal{U}^{2 I}_aV^a-  \left(\widebar\psi_1 \zeta_\Ibar \delta^{I\Ibar} -  \widebar\psi^2\zeta^I\right)\,,\\
    \widebar E^\Ibar \equiv   \;\mathcal{U}^{2 \Ibar}&=  \mathcal{U}^{2 \Ibar}_a V^a + \widebar\psi_1\zeta_I \delta^{I\Ibar}+ \widebar\psi^2\zeta^\Ibar \,, \\\label{imibar}
    i\,d\widebar M^\Ibar \equiv   \;   \mathcal{U}^{1 \Ibar}&= \mathcal{U}^{1 \Ibar}_aV^a + \widebar\psi^1\zeta^\Ibar-\widebar\psi_2\zeta_I \delta^{I\Ibar} \,.
\end{align}
\end{subequations}
{In particular, we remark that  the superspace parametrizations \eqref{imi}, \eqref{imibar} identify the quantities $M^I, \widebar M^{\Ibar}$, appearing in \eqref{defH} and \eqref{defHb} and required for consistency to satisfy \eqref{dm}, with the scalars of the parent hypermultiplets that have been dualized into tensors.}

\subsection{Supersymmetry transformation laws in superspace}

In the geometric approach to supersymmetric theories in superspace, it is straightforward to derive the supersymmetry transformation laws of the fields, since they amount to diffeomorphisms in the  odd directions of superspace which, in the case of rigid supersymmetry, reduce to super-translations (that is rigid translations of the odd coordinates
 $\theta^A_\alpha \to \theta^A_\alpha+\epsilon^A_\alpha$ along the odd directions $\theta^A_\alpha$, with constant Grassman{n} parameter $\epsilon^A_\alpha$,  generated by the supercharges $Q_A^\alpha$). As such, they can be accounted for through the corresponding \emph{Lie derivative} of the superfields. If we generically denote by $\Phi(x,\theta)$ a given superfield, then its supersymmetry transformation law in superspace is
\begin{align}
\delta_\epsilon \Phi=    \ell_\epsilon \Phi= d\iota_\epsilon (\Phi) + \iota_\epsilon (d\Phi)\,.
\end{align}
Here $\iota_\epsilon$ denotes contraction along the tangent-space vector (see Appendix \ref{superalgebraapp})$$\epsilon\equiv \epsilon_{{ \alpha} A} Q^{{ \alpha} A }+ \epsilon^{{ \alpha} A} Q^{ \alpha}_{ A }\,,$$ 
such that
\begin{align}
    \iota_\epsilon(\psi^A)= \epsilon^A\,,\quad \iota_\epsilon(\psi_A)= \epsilon_A\,.
\end{align}
 Using \eqref{def}-\eqref{defb} and \eqref{param}-\eqref{paramb}, we find
\begin{subequations}
\label{supersusy}
\begin{align}
   \delta_\epsilon L^I=& \iota_\epsilon(E^I)= \widebar\epsilon^1\zeta^{I}+\widebar\epsilon_2 \zeta_{\widebar I}\delta^{I\widebar I}  \,,  \\  
   \delta_\epsilon B^I=& \iota_\epsilon(dB^I)=\iota_\epsilon\left\{ -\left[8L^I\widebar\psi_1\gamma_a\psi^2  - 4i M^I \left(\widebar\psi_1\gamma_a\psi^1-\widebar\psi_2\gamma_a\psi^2\right)\right]V^a+H^I\right\}\nonumber\\
   =&-\left[8L^I\left(\widebar\epsilon_1\gamma_a\psi^2+\widebar\epsilon^2\gamma_a\psi_1\right)  - 4i M^I \left(\widebar\epsilon_1\gamma_a\psi^1+\widebar\epsilon^1\gamma_a\psi_1-\widebar\epsilon_2\gamma_a\psi^2-\widebar\epsilon^2\gamma_a\psi_2\right)\right]V^a \nonumber\\&-2 i  \left(\widebar\epsilon_1   \gamma_{ab}\zeta_{\widebar J}\delta^{I\widebar J}+\widebar\epsilon^2\gamma_{ab}\zeta^{I} \right)V^aV^b  \,, \\
   \delta_\epsilon \zeta_{\Ibar}=& \iota_\epsilon(d\zeta_{\Ibar})=    \delta_{I\Ibar}\left(i E^I_a \gamma^a\epsilon^2 -\frac 32 \,i h^I_a \gamma^a\epsilon^1\right) \,, \\
    \delta_\epsilon \zeta^I=& \iota_\epsilon(d\zeta^I)=   i E^I_a \gamma^a\epsilon_1 + \frac 32\,i h^I_a \gamma^a\epsilon_2 \,,
  \end{align} 
\end{subequations}
and 
\begin{subequations}
\label{supersusyb}
\begin{align}
   \delta_\epsilon \widebar L^\Ibar=& \iota_\epsilon(\widebar E^\Ibar)= \widebar\epsilon_1\zeta_{I}\delta^{I\widebar I} +\widebar\epsilon^2 \zeta^{\widebar I} \,,  \\  
   \delta_\epsilon \widebar B^\Ibar=& \iota_\epsilon(d\widebar B^\Ibar)=\iota_\epsilon\left\{ \left[8\widebar L^\Ibar\widebar\psi^1\gamma_a\psi_2  + 4i \widebar M^\Ibar \left(\widebar\psi_1\gamma_a\psi^1-\widebar\psi_2\gamma_a\psi^2\right)\right]V^a+\widebar H^\Ibar\right\}\nonumber\\
   =&\left[8\widebar L^\Ibar\left(\widebar\epsilon^1\gamma_a\psi_2+\widebar\epsilon_2\gamma_a\psi^1\right)  + 4i\widebar M^\Ibar \left(\widebar\epsilon_1\gamma_a\psi^1+\widebar\epsilon^1\gamma_a\psi_1-\widebar\epsilon_2\gamma_a\psi^2-\widebar\epsilon^2\gamma_a\psi_2\right)\right]V^a \nonumber\\&+2 i  \left(\widebar\epsilon^1   \gamma_{ab}\zeta^{\widebar I}+\widebar\epsilon_2\gamma_{ab}\zeta_{I} \delta^{I\widebar I}\right)V^aV^b  \,, \\
   \delta_\epsilon \zeta^{\Ibar}=& \iota_\epsilon(d\zeta^{\Ibar})=    i \widebar E^\Ibar_a \gamma^a\epsilon_2 -\frac 32 \,i\widebar  h^\Ibar_a \gamma^a\epsilon_1 \,, \\
    \delta_\epsilon \zeta_I=& \iota_\epsilon(d\zeta_I)=  \delta_{I\Ibar}\left( i \widebar E^\Ibar_a \gamma^a\epsilon^1 + \frac 32\,i \bar h^\Ibar_a \gamma^a\epsilon^2 \right)\,.
  \end{align} 
\end{subequations}

\subsection{The superspace Lagrangian}

The supersymmetry invariant 4-form  Lagrangian in superspace, constructed geometrically with the rheonomic approach \cite{Castellani:1991eu}, 
reads
\begin{align}
    \mathcal{L}_{\text{superspace}}& = a_1 \delta_{I\Jbar}\left[\tilde{E}^{I|a}\left(\widebar E^\Jbar-\widebar\psi_1\zeta_J\delta^{J\Jbar}-\widebar\psi^2\zeta^\Jbar\right) + \tilde{\widebar E}^{\Jbar|a}\left( E^I-\widebar\psi^1\zeta^I-\widebar\psi_2\zeta_\Ibar \delta^{I\Ibar}\right)\right]V^bV^cV^d\epsilon_{abcd} \nonumber \\
    &+ a_2 {\tilde{\bar h}^\Ibar_c \delta_{I\Ibar}}\left[H^I+2\,i\, \left(\widebar\psi_1\gamma_{ab}\zeta_\Jbar \delta^{I\Jbar}+\widebar\psi^2\gamma_{ab}\zeta^I\right) V^aV^b\right]V^c \nonumber \\
    &+ a_2{\tilde h^I_c \delta_{ I\Ibar}}\left[\widebar H^\Ibar -2\,i\,\left(\widebar\psi^1\gamma_{ab}\zeta^\Ibar +\widebar\psi_2\gamma_{ab}\zeta_J\delta^{J\Ibar}\right)   V^aV^b\right]V^c \nonumber \\ 
    &-\frac 14 \left(a_1\delta_{I\Jbar}\tilde{E}^{I|\ell}\tilde{\widebar{E}}^{\Jbar }_\ell + {a_2\tilde{h}^{I|\ell}\tilde{\bar h}^{\Jbar }_\ell \delta_{I\Jbar}} \right)V^aV^bV^cV^d\epsilon_{abcd}
    \nonumber \\
    &+ \frac{a_3}{4} \left( \bar\zeta_I\gamma^a\mathcal{D}\zeta^I +\bar\zeta_\Jbar\gamma^a\mathcal{D}\zeta^\Jbar
    +\bar\zeta^I\gamma^a\mathcal{D}\zeta_I +\bar\zeta^\Jbar\gamma^a\mathcal{D}\zeta_\Jbar\right)V^bV^cV^d\epsilon_{abcd} \nonumber \\
    &+ b_1 E^I \left(\bar\zeta_I\gamma_{ab} \psi_1 -\delta_{I\Jbar}\bar\zeta^\Jbar\gamma_{ab} \psi^2\right)  V^aV^b  
    +\widebar b_1 \widebar E^\Ibar \left(\bar\zeta_\Ibar\gamma_{ab} \psi_2 -\delta_{J\Ibar}\bar\zeta^J\gamma_{ab} \psi^1\right)  V^aV^b \nonumber \\
    & + i b_3\, H^I \left( \widebar\psi^1\zeta^\Ibar\delta_{I\Ibar} -\widebar\psi_2 \zeta_I \right) + i\widebar b_3\,\widebar H^\Ibar \left(\widebar\psi_1\zeta_\Ibar -\widebar\psi^2 \zeta^I\delta_{I\Ibar} \right)  \nonumber \\
    &+ c_1\left(\bar\zeta_I\gamma_a \zeta^I - \bar\zeta_\Ibar\gamma_a \zeta^\Ibar\right)\left(\widebar\psi_1\gamma_b \psi^1 - \widebar\psi_2\gamma_b \psi^2\right)V^a V^b \nonumber\\
    &+ c_2 \left(\bar\zeta_I\zeta_\Ibar \widebar\psi_1\gamma_{ab}\psi_2-\bar\zeta^I\zeta^\Ibar \widebar\psi^1\gamma_{ab}\psi^2\right)V^aV^b \nonumber\\
    &+ c_4 \left(L^I \widebar E^{\Ibar}-\widebar L^\Ibar E^I \right)\delta_{I\Ibar}\left(\widebar\psi_1\gamma_a\psi^1-\widebar\psi_2\gamma_a\psi^2\right)V^a \nonumber\\
    &+ c_5\left( M^I \mathcal{U}^{1\Ibar}-\widebar M^\Ibar \mathcal{U}^{2I}  \right) \delta_{I\Ibar} \left(\widebar\psi_1\gamma_a\psi^1-\widebar\psi_2\gamma_a\psi^2\right)V^a \,,
\end{align}
where
\begin{equation}\begin{split}
    & a_2=-\frac 94 \,a_1\,,\quad a_3= -2ia_1\,,\quad  b_1= \widebar b_1 = 3ia_1 \,,\quad b_3 = \widebar b_3 =-\frac 32 i a_1\,,\\
    & c_1=c_2=3 i a_1\,,\quad \tilde c_1= \tilde c_2=-3 i a_1\,,\quad c_4= 3 a_1\,,\quad c_5= -3ia_1 \,,
\end{split}\end{equation}
so that
\begin{align}
    \mathcal{L}_{\text{superspace}}& = a_1 \Bigl\{\delta_{I\Jbar}\left[\tilde{E}^{I|a}\left(\widebar E^\Jbar-\widebar\psi_1\zeta_J\delta^{J\Jbar}-\widebar\psi^2\zeta^\Jbar\right) + \tilde{\widebar E}^{\Jbar|a}\left( E^I-\widebar\psi^1\zeta^I-\widebar\psi_2\zeta_\Ibar \delta^{I\Ibar}\right)\right]V^bV^cV^d\epsilon_{abcd} \nonumber \\
    &-\frac 94{\tilde{\bar h}^\Ibar_c \delta_{I\Ibar}}\left[H^I+2\,i\, \left(\widebar\psi_1\gamma_{ab}\zeta_\Jbar \delta^{I\Jbar}+\widebar\psi^2\gamma_{ab}\zeta^I\right) V^aV^b\right]V^c \nonumber \\
    &-\frac 94{\tilde h^I_c \delta_{ I\Ibar}}\left[\widebar H^\Ibar -2\,i\,\left(\widebar\psi^1\gamma_{ab}\zeta^\Ibar +\widebar\psi_2\gamma_{ab}\zeta_J\delta^{J\Ibar}\right)   V^aV^b\right]V^c \nonumber \\ 
    &-\frac 14 \delta_{I\Jbar}\left(\delta_{I\Jbar}\tilde{E}^{I|\ell}\tilde{\widebar{E}}^{\Jbar }_\ell -\frac 94 {\tilde{h}^{I|\ell}\tilde{\bar h}^{\Jbar }_\ell } \right)V^aV^bV^cV^d\epsilon_{abcd}
    \nonumber \\
    &- \frac{i}{2} \left( \bar\zeta_I\gamma^a\mathcal{D}\zeta^I +\bar\zeta_\Jbar\gamma^a\mathcal{D}\zeta^\Jbar
    +\bar\zeta^I\gamma^a\mathcal{D}\zeta_I +\bar\zeta^\Jbar\gamma^a\mathcal{D}\zeta_\Jbar\right)V^bV^cV^d\epsilon_{abcd} \nonumber \\
    &+3i \left[E^I \left(\bar\zeta_I\gamma_{ab} \psi_1 -\delta_{I\Jbar}\bar\zeta^\Jbar\gamma_{ab} \psi^2\right)  - \widebar E^\Ibar \left(\delta_{J\Ibar}\bar\zeta^J\gamma_{ab} \psi^1 -\bar\zeta_\Ibar\gamma_{ab} \psi_2 \right)\right]  V^aV^b \nonumber \\
    & + \frac 32\, H^I \left( \widebar\psi^1\zeta^\Ibar\delta_{I\Ibar} -\widebar\psi_2 \zeta_I \right) + \frac 32\,\widebar H^\Ibar \left(\widebar\psi_1\zeta_\Ibar -\widebar\psi^2 \zeta^I\delta_{I\Ibar} \right)  \nonumber \\
    &+ 3i\left(\bar\zeta_I\gamma_a \zeta^I - \bar\zeta_\Ibar\gamma_a \zeta^\Ibar\right)\left(\widebar\psi_1\gamma_b \psi^1 - \widebar\psi_2\gamma_b \psi^2\right)V^a V^b \nonumber\\
 &+ 3i \left(\bar\zeta_I\zeta_\Ibar \widebar\psi_1\gamma_{ab}\psi_2-\bar\zeta^I\zeta^\Ibar \widebar\psi^1\gamma_{ab}\psi^2\right)V^aV^b \nonumber\\
      &+ 3 \delta_{I\Ibar}\left(L^I \widebar E^{\Ibar}-   \widebar L^\Ibar E^I \right)\left(\widebar\psi_1\gamma_a\psi^1-\widebar\psi_2\gamma_a\psi^2\right)V^a \nonumber\\
    &-3i\,\delta_{I\Ibar}\left( M^I 
    \mathcal{U}^{1\Ibar}-\widebar M^\Ibar \mathcal{U}^{2I} \right) \left(\widebar\psi_1\gamma_a\psi^1-\widebar\psi_2\gamma_a\psi^2\right)V^a\Bigr\} \,.
\label{superlag}
\end{align}
We are now going to discuss the Euler-Lagrange equations in superspace, derived from the above Lagrangian.

\subsubsection*{Field equations}

The {spacetime} field equations are given by the internal components (that is   {those} along products of the vierbein $V^a$ only) of the field equations in superspace, while the other components have to vanish identically. {This is a necessary condition for the theory in superspace to be equivalent to the one on spacetime, and they do  indeed vanish identically, upon using the parametrizations \eqref{param}-\eqref{paramb}.}

The field equations of the auxiliary fields $\tilde E^I_a, \tilde h^I_a$ and of their complex conjugates identify them on-shell with the supercovariant field strengths appearing in the parametrizations \eqref{param}-\eqref{paramb}:
\begin{subequations}\label{eqaux}
\begin{align}
\frac{\delta \mathcal{L}} {\delta \tilde E^I_a}=0:\quad \tilde E^I_a=  E^I_a\,,&\qquad   \frac{\delta \mathcal{L}} {\delta \tilde {\widebar E}^\Ibar_a}=0:\quad\tilde {\widebar E}^\Ibar_a=  \widebar E^\Ibar_a\,,\\
\frac{\delta \mathcal{L}} {\delta \tilde h^I_a}=0:\quad \tilde h^I_a=h^I_a= \frac 16 \epsilon_{abcd}H^{I | bcd}\,,  &\qquad \frac{\delta \mathcal{L}} {\delta \tilde {\bar h}^\Ibar_a}=0:\quad \tilde {\bar h}^\Ibar_a=\bar h^\Ibar_a= \frac 16 \epsilon_{abcd}\widebar H^{\Ibar |bcd}\,.
\end{align}
\end{subequations}
Then, using \eqref{eqaux}, the field equations for the other bosonic fields read
\begin{subequations}
\label{eqB}
\begin{align}
\frac{\delta \mathcal{L}} {\delta L^I}=0:\quad \partial^a\partial_a \widebar L^\Ibar=0\,,&\qquad   \frac{\delta \mathcal{L}} {\delta  {\widebar L}^\Ibar}=0:\quad
\partial^a\partial_a L^I=0\,;\\
\frac{\delta \mathcal{L}} {\delta B^I}=0:\quad \partial^a\partial_{[a}\widebar B^\Ibar_{bc]} =0\,,&\qquad   \frac{\delta \mathcal{L}} {\delta  {\widebar B}^\Ibar}=0:\quad
\partial^a\partial_{[a} B^I_{bc]}=0\,.
\end{align}
\end{subequations}
Finally, the field equations of the spinors are
\begin{align}\label{eqF}
\slashed\partial\zeta_I=0\,,\quad \slashed\partial\zeta^\Ibar=0 \,,\quad    \slashed\partial\zeta^I=0\,,\quad \slashed\partial\zeta_\Ibar=0\,.
\end{align}
Eqs. \eqref{eqB} and \eqref{eqF} are the field equations of the set of free double-tensor multiplets. 
All the other contributions to the variations in superspace, but also the variations of the superspace Lagrangian with respect to the scalars $M^I,\widebar M^\Ibar$ that have been dualized into the 2-forms $B^I,\widebar B^\Ibar$, vanish identically {thanks to} \eqref{param}-\eqref{paramb}.

Anyhow, these scalars  satisfy the harmonic  equations:
\begin{align} \label{327}\partial_a\partial^a M^I=0\,,\quad 
\partial_a\partial^a \widebar M^\Ibar=0\,.\end{align}
This is due to the fact that, as pointed out below, in eqs. \eqref{relhM}, they are on-shell related to the tensors $B^I_{ab}$ by
\begin{align}\label{MB}
  \partial_a M^I = -{\frac14} i \epsilon_{abcd}\partial^b B^{I cd}  \,,\quad \partial_a \widebar M^\Ibar = {\frac14} i \epsilon_{abcd}\partial^b \widebar B^{\Ibar cd} \,,
\end{align}
so that the conditions \eqref{327} hold identically:
\begin{align}
 \partial^a \partial_a M^I = -{\frac14} i \epsilon_{abcd}\partial^a\partial^b B^{I cd}  \equiv 0\,,
\end{align}
and the same for the complex conjugates.

\subsection{Remarks emerging from the superspace analysis}\label{remarks}

From the closure of the Bianchi identities in superspace, we find, in particular,
\begin{subequations}
\label{relhM}
\begin{align}
    h^I_{a} & = \frac 23 \,i \partial_a M^I=\frac 16 \epsilon_{abcd}H^{I|bcd} \,, \label{relhq} \\
    \widebar{h}^\Ibar_{a} & = -\frac 23 \,i \partial_a\widebar{M}^\Ibar = \frac 16 \epsilon_{abcd}\widebar{H}^{\Ibar|bcd} \,, \label{relhqbar}
\end{align}
\end{subequations}
together with the following superspace parametrization for the quantities $dM^I$ appearing in the superspace definition of the 3-form field strengths:
\begin{subequations}
\label{dMU-h}
\begin{align}
    dM^I = - i\, \mathcal{U}^{2I} &= -\frac 32 \,i\, h^I_a V^a +\,i \left( \widebar\psi_1\zeta_{\Jbar}\delta^{J \Jbar} -\widebar\psi^2\zeta^J
    \right) \,, \\
    d\widebar M^{\Ibar} = -i \,\mathcal{U}^{1\Ibar} & = \frac 32 \,i\, \bar h^{\Ibar}_a V^a -\,i \left( \widebar\psi^1 \zeta^{\Ibar} - \widebar \psi_2 \zeta_J \delta^{\Ibar J} \right) \,.
\end{align}
\end{subequations}
Furthermore, we get
\begin{subequations}
\label{dpsih}
\begin{align}
    {\buildrel \psi  \over \nabla }h^I_a& = - \frac 23 \left(\widebar\psi_1\partial_a\zeta_\Ibar \delta^{I \Ibar}- \widebar\psi^2\partial_a\zeta^I \right) \,, \\
    {\buildrel \psi  \over \nabla }\bar h^\Ibar_a& = - \frac 23 \left( \widebar\psi^1\partial_a\zeta^\Ibar - \widebar\psi_2\partial_a\zeta_I\delta^{I \Ibar} \right) \,.
\end{align}
\end{subequations}
The above relations identify the $M^I$'s (and $\widebar M^\Ibar$'s) with the scalar isometries that have been dualized into tensor fields. Surprisingly, they need necessarily to be included in the definition of the super-field strengths, for the consistency of the theory. 
Moreover, from the fact that  {on-shell}
\begin{equation}
 h^I_\mu = \frac 16 \epsilon_{\mu\nu\rho\sigma} H^{I \nu\rho\sigma}= {\frac16} \epsilon_{\mu\nu\rho\sigma} \partial^\nu B^{I \rho\sigma},\label{hH}   
\end{equation} 
we get the harmonic conditions \eqref{327},
which are the \emph{free} equations of motion for the set of scalars $M^I,\widebar M^\Ibar$. Therefore, despite the surprising fact that the consistency of the off-shell (self-interacting) theory requires the introduction of extra \emph{unphysical} d.o.f. -- associated with the scalars that have been dualized -- the latter satisfy on-shell the field equations of free scalar fields. This follows from the algebraic duality relation \eqref{hH} and, as such, it holds true also in more complicated models, where the double-tensor multiplets are coupled to  {other supersymmetric multiplets}. As a consequence, the scalars $M^I$ are fully decoupled, on-shell, from the physical sector.\footnote{This is reminiscent of what happens in chiral ($4n+2$)-dimensional supersymmetric models, where the 
on-shell matching of degrees of freedom requires the ($2n+1$)-form field strengths of tensor multiplets to be  self-dual.   An action principle,  implementing the self-duality constraint dynamically, was found by A. Sen 
\cite{Sen:2015nph}, at the cost of introducing off-shell new, unphysical d.o.f. (in that case, extra $2n$-indices antisymmetric tensor fields), satisfying on-shell harmonic  equations, and  then fully decoupled from the physical spectrum. We will further elaborate on this in the Conclusions.}

Let us now observe that, given the relations \eqref{relhM} and \eqref{dMU-h}, we have in fact two different expressions for the spinorial derivative of $h^I_a, \bar h^\Ibar_a$: one is given by \eqref{dpsih}, the other is obtained by evaluating the spinorial derivative of $H^{I|abc}$ and $\widebar H^{\Ibar|abc}$, which read
\begin{subequations}
\label{dpsiH}
\begin{align}
  & {\buildrel \psi  \over \nabla } H^I_{abc} = -2i\left( \widebar\psi_1 \gamma_{[ab} \partial_{c]} \zeta_{\Ibar} \delta^{I \Ibar} + \widebar \psi^2 \gamma_{[ab} \partial_{c]} \zeta^I  \right)\,,  \\
  & {\buildrel \psi  \over \nabla } \widebar H^{\Ibar}_{abc} = 2i \left( \widebar\psi^1 \gamma_{[ab} \partial_{c]} \zeta^{\Ibar} + \widebar \psi_2 \gamma_{[ab} \partial_{c]} \zeta_I  \delta^{I \Ibar}\right) \,. 
\end{align}
\end{subequations}
Using \eqref{dpsiH} in \eqref{relhM}, instead we get 
\begin{subequations}
\label{dpsihH}
\begin{align}
  {\buildrel \psi  \over \nabla }h^I_a &= \frac 16 \epsilon_{abcd} {\buildrel \psi  \over \nabla }H^{I|bcd} = {- \frac{i}{3} \epsilon_{abcd} \left( \widebar\psi_1 \gamma^{bc} \partial^{d} \zeta_{\Ibar} \delta^{I \Ibar} + \widebar \psi^2 \gamma^{bc} \partial^{d} \zeta^I \right)} \nonumber \\
  &= \frac23 \left( \widebar\psi_1 \gamma_{ad} \partial^{d} \zeta_{\Ibar} \delta^{I \Ibar} {-} \widebar \psi^2 \gamma_{ad} \partial^{d} \zeta^I \right) \,, \\
  {\buildrel \psi  \over \nabla }\bar h^\Ibar_a &= \frac 16 \epsilon_{abcd} {\buildrel \psi \over \nabla }\widebar H^{\Ibar|bcd} = {\frac{i}{3} \epsilon_{abcd} \left( \widebar\psi^1 \gamma^{bc} \partial^{d} \zeta^{\Ibar} + \widebar \psi_2 \gamma^{bc} \partial^{d} \zeta_I  \delta^{I \Ibar}\right)} \nonumber \\
  &= -\frac23 \left({-} \widebar\psi^1 \gamma_{ad} \partial^{d} \zeta^{\Ibar} + \widebar \psi_2 \gamma_{ad} \partial^{d} \zeta_I  \delta^{I \Ibar}\right) \,.
\end{align}
\end{subequations}
Requiring equivalence of \eqref{dpsih} with \eqref{dpsihH}
{implies}
\begin{subequations}
\begin{align}
\widebar\psi_1\partial_a\zeta_\Ibar \delta^{I \Ibar}- \widebar\psi^2\partial_a\zeta^I  &= - \left( \widebar\psi_1 \gamma_{ad} \partial^{d} \zeta_{\Ibar} \delta^{I \Ibar} {-} \widebar \psi^2 \gamma_{ad} \partial^{d} \zeta^I \right) \,, \\
\widebar\psi^1\partial_a\zeta^\Ibar - \widebar\psi_2\partial_a\zeta_I\delta^{I \Ibar}  &= {-}\widebar\psi^1 \gamma_{ad} \partial^{d} \zeta^{\Ibar} + \widebar \psi_2 \gamma_{ad} \partial^{d} \zeta_I  \delta^{I \Ibar} \,,
\end{align}
\end{subequations}
which are satisfied if
\begin{align*}
\slashed\partial\zeta_I=0\,,\quad \slashed\partial\zeta^\Ibar=0 \,,\quad    \slashed\partial\zeta^I=0\,,\quad \slashed\partial\zeta_\Ibar=0\,.
\end{align*}
The latter are the field equations for the spinors, eqs. \eqref{eqF}, and this shows, already at the level of the  Bianchi identities of the bosonic fields in superspace, that the double-tensor multiplet is an on-shell multiplet.
The same result is retrieved by inspecting the Bianchi identities of the spinors themselves in superspace. Here, the on-shell condition is found in the $2\psi$-sector of the superspace Bianchi identities.

\section{Spacetime description of the model}\label{spacetimemodel}

We shall now derive the supersymmetry transformation laws and the supersymmetry invariant Lagrangian of the double-tensor multiplets in the spacetime component approach. We will see that the relevant results obtained above in superspace are also retrieved as physical conditions when we restrict ourselves to spacetime.

\subsection{Supersymmetry transformation laws on spacetime}

In rigid, flat superspace the supervielbein reads
\begin{subequations}\begin{align}
    V^a &= V^a_\mu dx^\mu + \frac{i}{2} \widebar{\theta}_A \gamma^a d\theta^A + \frac{i}{2} \widebar{\theta}^A \gamma^a d\theta_A \,, \\
    \psi_A &= d\theta_{A} \,, \\ 
    \psi^A &= d\theta^{A} \,,
\end{align}\end{subequations}
{where $V^a_\mu$ is the spacetime vierbein, and $\theta_A,\theta^A$ are the left- and right-chirality projections of the Grassmann-odd coordinates of superspace.}
It is then straightforward to derive the supersymmetry transformation laws in spacetime, by projecting on spacetime eqs. \eqref{supersusy}-\eqref{supersusyb}. They
read
\begin{subequations}
\label{susytransf}
\begin{align} 
\delta_\epsilon L^I &= \widebar\epsilon^1\zeta^I+\widebar\epsilon_2\zeta_\Ibar \delta^{I\Ibar}\,,\\ \delta_\epsilon B^I_{\mu\nu} &= - {2} i \left(\widebar\epsilon_1\gamma_{\mu\nu}\zeta_\Ibar\delta^{I\Ibar}+\widebar\epsilon^2\gamma_{\mu\nu}\zeta^I\right)\,,\\
\label{zetal}  \delta_\epsilon \zeta_\Ibar&= i\,\partial_\mu L^I \gamma^\mu \epsilon^2\delta_{I\Ibar} -\frac 32 \,i h^I_\mu \gamma^\mu \epsilon^1 \delta_{I\Ibar} \,, \\
    \delta_\epsilon \zeta^I&= i\,\partial_\mu L^I \gamma^\mu \epsilon_1 +\frac 32 \,i h^I_\mu \gamma^\mu \epsilon_2\,,
\label{zetar}    \end{align}
\end{subequations}    
their complex conjugates being
\begin{subequations}
\label{susytransfbar}
\begin{align}
    \delta_\epsilon \widebar L^\Ibar &= \widebar\epsilon_1\zeta_I\delta^{I\Ibar}+\widebar\epsilon^2\zeta^\Ibar \,,\\
    \delta_\epsilon \widebar B^\Ibar_{\mu\nu} &= {2} i \left(\widebar\epsilon^1\gamma_{\mu\nu}\zeta^\Ibar+\widebar\epsilon_2\gamma_{\mu\nu}\zeta_I\delta^{I\Ibar}\right)\,,\\
    \delta_\epsilon \zeta^\Ibar&= i\,\partial_\mu \widebar L^\Ibar \gamma^\mu \epsilon_2 -\frac 32 \,i\bar h^\Ibar_\mu \gamma^\mu \epsilon_1\,,\\
    \delta_\epsilon \zeta_I &= i\,\partial_\mu \widebar L^\Ibar \gamma^\mu \epsilon^1\delta_{I\Ibar} +\frac 32 \,i \bar h^\Ibar_\mu \gamma^\mu \epsilon^2 \delta_{I\Ibar} \,. 
\end{align}
\end{subequations}
This result coincides with the direct computation obtained by acting with supersymmetry transformations, generated by the supercharges $Q_A,Q^A$, on the spacetime restriction of the fields, generically denoted by $\Phi(x)$:
\begin{align}
\delta_\epsilon \Phi(x) = \left[\left(\epsilon_\alpha^A Q^\alpha_A +\epsilon^\alpha_A Q_\alpha^A\right),\Phi(x)\right]
\,.
\end{align}

\subsubsection*{Supersymmetry invariance } 

Let us inspect here  {the closure of the supersymmetry algebra on} the double-tensor multiplets restricted to spacetime.
This amounts to  check  under which conditions the commutator of two supersymmetry transformations, with spinor-parameters $\epsilon,\sigma$ respectively, does reproduce the supersymmetry algebra \eqref{superalgapp}, with $p_\mu=\partial_\mu$ and vanishing central charges {(since we are considering massless representations}):
\begin{equation}\begin{split}
    \label{commu}
    \left[\delta_\epsilon,\delta_\sigma\right]&=\left[\epsilon_{A\alpha} Q^{A\alpha},\sigma^{B}_\beta Q_{B}^\beta\right] + \left[\epsilon^{A}_{\hat\alpha} Q^{\hat\alpha}_{A},\sigma_{\hat \beta B} Q^{\hat\beta B}\right] \\
    &= -\epsilon_{A\alpha}\left\{Q^{\alpha A},Q^\beta_{B}\right\}\sigma^B_{\beta}-\epsilon^{A}_\alpha\left\{Q^{\alpha}_A,Q^{\beta B}\right\}\sigma_{B\beta} \\
    &=-i \left(\widebar{\epsilon}_A\gamma^\mu\sigma^A+\widebar{\epsilon}^A\gamma^\mu\sigma_A \right)p_\mu \\ &=i\left(\widebar\sigma_A\gamma^\mu\epsilon^A+\widebar\sigma^A\gamma^\mu\epsilon_A \right)\partial_\mu \,.
\end{split}\end{equation}
Applying the commutator of two supersymmetry transformations, eq.  \eqref{commu}, to the holomorphic  scalars $L^I$, we get
\begin{align}
    \left[ \delta_\epsilon,\delta_\sigma \right]L^I &= i\left(\widebar\sigma_A\gamma^\mu\epsilon^A+\widebar\sigma^A\gamma^\mu\epsilon_A \right)\partial_\mu L^I  \,,
 \end{align}
consistently with \eqref{commu}.

On the other hand, 
the action of \eqref{commu} on the antisymmetric tensors $B^I_{\mu\nu}$ gives
\begin{equation}\begin{split}
    \left[\delta_\epsilon,\delta_\sigma\right]B^I_{\mu\nu}&=3i\left(\widebar\sigma_A\gamma^\rho\epsilon^A+\widebar\sigma^A\gamma^\rho\epsilon_A \right)\partial_{[\rho} B^I_{\mu\nu]}  \\
    &+{4}\partial_{[\mu}\left[L^I\left(\widebar\sigma_1\gamma_{\nu]}\epsilon^2+\widebar\sigma^2\gamma_{\nu]}\epsilon_1 \right)\right]- {\frac12}h^I_{[\mu}\left(\widebar\sigma_A\gamma_{\nu]}\epsilon^A-\widebar\sigma^A\gamma_{\nu]}\epsilon_A \right)\,,\label{d2tens}
\end{split}\end{equation}
which is not in the form \eqref{commu} required for closure, {in particular because of the term in $h^I_\mu$, which appears in the supersymmetry transformation laws \eqref{zetal}, \eqref{zetar} and is related to the tensors field strengths by} $h^I_\mu \,\propto\,  \epsilon_{\mu\nu\rho\sigma}\partial^\nu B^{\rho\sigma}$. However, 
if we use {instead for $h^I_\mu$} the expression \eqref{relhM}, that is 
$h^I_\mu=\frac{2}{3}\,i \,\partial_\mu M^I$,
{eq. \eqref{d2tens}} can be rewritten in the form \eqref{commu} \emph{up to a gauge transformation}
\begin{equation}
    B_{\mu\nu} \to B_{\mu\nu} +\partial_{[\mu}X^I_{\nu]}\,,
\end{equation}
that is
\begin{align}\label{delta2b}
\left[\delta_\epsilon,\delta_\sigma\right]B^I_{\mu\nu}&=i\left(\widebar\sigma_A\gamma^\rho\epsilon^A+\widebar\sigma^A\gamma^\rho\epsilon_A \right)\partial_\rho B^I_{\mu\nu} +\partial_{[\mu}X^I_{\nu]}\,,
\end{align}
since in this case we can define:
\begin{equation}\begin{split}
    X^I_\nu &\equiv 
    2i \left( \widebar\sigma_A\gamma^\rho\epsilon^A + \widebar\sigma^A\gamma^\rho\epsilon_A \right) B^I_{\nu\rho} \\
    &+ {4} \left[ L^I \left( \widebar\sigma_1\gamma_{\nu}\epsilon^2 + \widebar\sigma^2\gamma_{\nu}\epsilon_1 \right) \right] - \frac {{i}}3 \,M^I \left( \widebar\sigma_A\gamma_{\nu}\epsilon^A -\widebar\sigma^A\gamma_{\nu}\epsilon_A \right) \,.
\end{split}\end{equation}

As far as the spinors of the multiplet are concerned, applying explicitly the commutator of two supersymmetry transformations to the spinors $\zeta^I$, and using
\begin{align}
    \delta_\sigma
    h^I_\mu&=-\frac i3\, \epsilon_{\mu\nu\rho\lambda}\left(\widebar\sigma_1\gamma^{\nu\rho}\partial^\lambda \zeta_\Ibar\,\delta^{I\Ibar}+\widebar\sigma^2\gamma^{\nu\rho}\partial^\lambda \zeta^I
    \right)\,,
\end{align}
we obtain
\begin{subequations}\begin{equation}\begin{split}
    \left[\delta_\epsilon,\delta_\sigma\right]\zeta^I &= i\left(\widebar\sigma_A\gamma^\mu\epsilon^A+\widebar\sigma^A\gamma^\mu\epsilon_A \right)\partial_\mu \zeta^I  \\
    &-\frac i2 \,\left(\widebar\sigma_1\gamma^\mu\epsilon^1+\widebar\sigma^1\gamma^\mu\epsilon_1 -\widebar\sigma_2\gamma^\mu\epsilon^2-\widebar\sigma^2\gamma^\mu\epsilon_2 \right)\gamma_\mu \slashed\partial\zeta^I \\
    &+i\,\left(\widebar\sigma_1\epsilon_2-\widebar\sigma_2\epsilon_1\right)\slashed\partial\zeta_\Ibar \,\delta^{I\Ibar} \,,
\end{split}\end{equation}
\begin{equation}\begin{split}
    \left[\delta_\epsilon,\delta_\sigma\right]\zeta_\Ibar &= i\left(\widebar\sigma_A\gamma^\mu\epsilon^A+\widebar\sigma^A\gamma^\mu\epsilon_A \right)\partial_\mu \zeta_\Ibar  \\
    &+\frac i2 \,\left(\widebar\sigma_1\gamma^\mu\epsilon^1+\widebar\sigma^1\gamma^\mu\epsilon_1 -\widebar\sigma_2\gamma^\mu\epsilon^2-\widebar\sigma^2\gamma^\mu\epsilon_2 \right)\gamma_\mu \slashed\partial\zeta_\Ibar \\
    &-i\,\left(\widebar\sigma_1\epsilon_2-\widebar\sigma_2\epsilon_1\right)\slashed\partial\zeta^I \,\delta_{I\Ibar} \,.
\end{split}\end{equation}\label{d2zeta}\end{subequations}
This reproduces the supersymmetry algebra \eqref{commu} only \emph{on-shell}, in which case the second and third lines of eqs. \eqref{d2zeta} vanish.

\subsection{The spacetime Lagrangian}

Projecting on spacetime the {superspace} Lagrangian \eqref{superlag}, going to second order on the scalar fields, that is setting  $\tilde E^I_\mu= E^I_\mu= \partial_\mu L^I$, and then choosing $a_1=-\frac 16$  to get the standard normalization of the scalars kinetic terms, we obtain
\begin{equation}\begin{split}
    \mathcal{L}_{\text{sp-t}}&=\Big[\left(\partial_\mu L^I \partial^\mu\widebar L^{\Jbar  }+\frac{9}{4}\tilde h^I_\mu \tilde{\bar h}^{\Jbar \mu}\right)\delta_{I\Jbar}+  \frac 38 \left(\tilde h^I_\sigma \widebar H^\Jbar_{\mu\nu\rho} +\tilde{\bar h}^\Jbar_\sigma H^I_{\mu\nu\rho}\right)\epsilon^{\mu\nu\rho\sigma}\delta_{I\Jbar} \\
    & -i \left(\bar\zeta_I\gamma^\mu\partial_\mu \zeta^I +
    \bar\zeta_\Ibar\gamma^\mu\partial_\mu \zeta^\Ibar  \right)\Big]d^4x  \,, \label{lst1}
\end{split}\end{equation}
corresponding to the hypermultiplets Lagrangian.
In \eqref{lst1}, just as in \eqref{superlag}, $\tilde h_\mu^I, \tilde {\bar h}_\mu^\Ibar$ are Lagrange multipliers implementing, at first order, the kinetic terms of the antisymmetric tensors. Upon variation of   \eqref{lst1}, we find
    \begin{align}
   \frac{\delta \mathcal{L}_{\text{sp-t}}}{\delta \tilde{\bar h}^\Ibar_\mu}=0 :\quad  \tilde h^I_\mu= \frac 16 \epsilon_{\mu\nu\rho\sigma}H^{I |\nu\rho\sigma } \,;\qquad   \frac{\delta \mathcal{L}_{\text{sp-t}}}{\delta \tilde{ h}^I_\mu}=0 :\quad \tilde{\bar h}^\Ibar_\mu= \frac 16 \epsilon_{\mu\nu\rho\sigma}\widebar H^{\Ibar| \nu\rho\sigma } \,, \label{lagmul}
    \end{align}
    which, plugged back into the spacetime Lagrangian, give
 \begin{align}
\mathcal{L}_{\text{sp-t}}
    &= \left[\left(E^I_\mu \widebar E^{\Jbar \mu}+\frac{3}{8}H^I_{\mu\nu\rho} \widebar H^{\Jbar \mu\nu\rho}\right)\delta_{I\Jbar} -i \left(\bar\zeta_I\gamma^\mu\partial_\mu \zeta^I + \bar\zeta_\Ibar\gamma^\mu\partial_\mu \zeta^\Ibar  \right)\right]d^4x \,.\label{lst2}
\end{align}
Using the supersymmetry transformation laws \eqref{susytransf} in eq. \eqref{lst2}, 
we find that the action is supersymmetry invariant off-shell upon identifying  the quantities $\tilde h^I_\mu$ and $\tilde {\bar h}^\Ibar_\mu$ in \eqref{lagmul} with the tensors  $h^I_\mu,\bar h^\Ibar_\mu$, appearing in the supersymmetry transformation laws of the spinors, that is by requiring:
\begin{align}
    h^I_\mu= \frac 16 \epsilon_{\mu\nu\rho\sigma} H^{I | \nu\rho\sigma } \,,\qquad  {\bar h}^\Ibar_\mu= \frac 16 \epsilon_{\mu\nu\rho\sigma}\widebar H^{\Ibar| \nu\rho\sigma } \,.
\end{align}
Correspondingly, the spacetime Lagrangian is invariant up to a boundary term:
\begin{align}
\label{totder}
 \delta_\epsilon \mathcal{L}_{\text{sp-t}} = &\partial_\mu \Big[ \partial_\nu L^I \left(\bar\zeta_I\epsilon_1\eta^{\mu\nu}-\delta_{I\Ibar}\bar\zeta^\Ibar \gamma^{\mu\nu}\epsilon^2\right) + \partial_\nu \bar L^\Ibar \left(\bar\zeta_\Ibar\epsilon_2\eta^{\mu\nu}-\delta_{I\Ibar}\bar\zeta^I \gamma^{\mu\nu}\epsilon^1\right) \nonumber\\
 &+\frac 32 h^I_\nu \left(\bar\zeta_I\gamma^{\mu\nu}\epsilon_2+\delta_{I\Ibar}\bar\zeta^\Ibar \epsilon^1\eta^{\mu\nu}\right)- \frac 32 \bar h^\Ibar_\nu \left(\bar\zeta_\Ibar\gamma^{\mu\nu}\epsilon_1+\delta_{I\Ibar}\bar\zeta^I \epsilon^2\eta^{\mu\nu}\right)\Bigr]\,.
\end{align}

\subsection{Remarks emerging from the spacetime analysis}

In the spacetime approach, the need to include the antisymmetric tensors $B^I_{\mu\nu}, \widebar B^\Ibar_{\mu\nu}$ \emph{together with} the scalars $M^I, \widebar M^\Ibar$ related to them via  Hodge-duality is subtle, and not so manifest as it is in the superspace approach. This condition is however required, and hidden in the double role of the quantities $h^I_\mu$ appearing in the supersymmetry transformations \eqref{zetal}-\eqref{zetar} of the spinors, as it is clarified in Section \ref{remarks} and summarized in eq. \eqref{relhM}:

\begin{itemize}
\item
At the Lagrangian level, a possible role  of the scalars $M^I$
is  not evident, since  the spacetime Lagrangian does not include them.
Moreover, the condition \eqref{totder} for the  invariance of the action  under supersymmetry requires $h^I_\mu$ to be related by Hodge-duality to the tensor field strengths $H^I_{\mu\nu\rho}$.
\item
The closure of the supersymmetry algebra on the tensors $B^I_{\mu\nu}$, eq. \eqref{delta2b}, requires instead $h^I_\mu$ to be  total derivatives, that is to satisfy  $h^I_\mu \propto \partial_\mu M^I$. 
\end{itemize}
We therefore obtain that the consistency of the supersymmetry algebra on the fields requires a double nature of the algebraic vectors $h_\mu^I$ in \eqref{zetal}-\eqref{zetar}: On the one hand, they have to be algebraically related to the tensor field strengths $H^I_{\mu\nu\rho}$ through \eqref{lagmul} to have invariance of the Lagrangian up to total derivative under the transformations \eqref{susytransf} and their complex conjugates; On the other hand, they have instead to be total derivatives $h^I_\mu\propto \partial_\mu M^I$, in order for the  transformations \eqref{susytransf} to close the supersymmetry algebra.

\section{Adding the cosmological constant}\label{cosmconstadd}

We would like to explore here the compatibility of our 
multiplets with the full centrally extended supersymmetry algebra. 
To this aim, we consider  {the embedding of   hypermultiplets  and double-tensor multiplets in} a rigid but curved supersymmetric background, that is a super  {AdS} background, with cosmological constant {$\Lambda= -\frac 3{\ell^2}$}. 

\subsection{The curved background}
The rigid, super AdS geometry is defined  by the Maurer-Cartan equations of the $OSp(2|4)$ superalgebra. In terms of 4-component gravitini $\Psi_A$, it reads
\begin{subequations}
\label{cosmconstbackLa}
\begin{align}
    \hat R^{ab} &\equiv d\hat\omega^{ab}+ \hat\omega^{ac}\hat\omega_c{}^b = \frac{1}{\ell^2}V^aV^b + \frac{1}{2\ell}\widebar\Psi_A\gamma^{ab}\Psi_B \delta_{AB} \,, \\
    T^a &\equiv \hat{\mathcal{D}}V^a - \frac i2\widebar\Psi_A\gamma^a\Psi_B \delta_{AB} = 0 \,, \\
    F &\equiv dA - \widebar\Psi_A\Psi_B\epsilon_{AB} = 0\,, \\
    \hat{\nabla}\Psi_A &\equiv \hat{\mathcal{D}}\Psi_A + \frac 1{2\ell} A \Psi_B \epsilon_{AB} =  + \frac{i}{2\ell}\gamma_a\Psi_A V^a  \,,
\end{align}
\end{subequations} 
where $\hat{\omega}^{ab}$ is the supertorsionless connection in the deformed background, $\hat{\mathcal{D}}$ is the corresponding Lorentz covariant derivative, and $\hat\nabla$ the full (Lorentz and gauge) covariant derivative. The  Bianchi identities
\begin{subequations}\begin{align}
    \hat{\mathcal{D}}\hat{R}^{ab} &= 0 \,,\\
    \hat{\mathcal{D}}T^a - \hat{R}^{ab}V_b + \frac i2 \hat\nabla\widebar\Psi_A\gamma^a\Psi_B \delta_{AB} - \frac i2 \widebar\Psi_A\gamma^a\hat\nabla\Psi_B \delta_{AB} &= 0  \,,\\
   dF + \hat\nabla\widebar\Psi_A\Psi_B\epsilon_{AB} - \widebar\Psi_A\hat\nabla\Psi_B\epsilon_{AB} &= 0 \,,\\
    \hat\nabla^2\Psi_A - \frac14\gamma_{ab}\hat{R}^{ab}\Psi_A - \frac{1}{2\ell}dA\Psi_B\epsilon_{AB} &= 0
\end{align}\end{subequations}
are identically satisfied, relying on the 3-$\Psi$ equation (Fierz identity):
\begin{equation}
\label{fierz!}\frac 1{4}\gamma_{ab}\Psi_A\widebar \Psi_B\gamma^{ab}\Psi_C \delta_{BC}-  \frac 1{2}\gamma_a \Psi_A\widebar \Psi_B\gamma^{a}\Psi_C\delta_{BC}=   \epsilon_{AB}\epsilon_{CD}\Psi_B\widebar \Psi_C \Psi_D \,.
\end{equation}
Here, the graviphoton $A$ is a background gauge field of the algebra $SO(2)\subset SU(2)_R$, which is the residual part of the R-symmetry $SU(2)_R\times U(1)$ preserved in the presence of the cosmological constant. 
We remark that, differently from the flat case, the  gravitino is minimally coupled with $A$ in the curved background, the AdS radius behaving as (the inverse of) an electric charge.
This implies that the electric-magnetic duality invariance of the graviphoton $A$ and of its Hodge-dual $\tilde A$, holding in the absence of gauge couplings, is now broken.\footnote{More precisely, $\tilde A_\mu$ is the potential whose field strength  $\partial_{[\mu}\tilde A_{\nu]}$ is Hodge-dual to $\partial_{[\mu}  A_{\nu]}$.}
Indeed, in \emph{flat} superspace the Maurer-Cartan equations of the background include $\tilde {\mathcal F}\equiv \star dA$, which is defined as
\begin{align}
    \tilde {\mathcal F}(=d \tilde A)=\widebar \Psi_A \gamma_5 \Psi_B \epsilon_{AB}\,,
\end{align}
and satisfies
\begin{align}
    d\tilde {\mathcal F}=0\,.
\end{align}
On the other hand, in the $OSp(2|4)$-invariant background, necessarily $\tilde {\mathcal F} \neq d\tilde A$, because in this case we get
\begin{align}\label{mag}
    d\tilde {\mathcal F}=-\frac i \ell\widebar \Psi_A \gamma_5 \gamma_a\Psi_B \epsilon_{AB}V^a \neq 0 \,.
\end{align}

\medskip
In terms of the left-handed and right-handed components of the gravitini, the supersymmetric background is
\begin{subequations}
\label{cosmconstbackchir}
\begin{align}
    \hat R^{ab} &\equiv d\hat\omega^{ab}+ \hat\omega^{ac}\hat\omega_c{}^b =  \frac{1}{\ell^2}V^aV^b  +\frac{1}{2\ell}\left(\widebar\psi_A\gamma^{ab}\psi_B \delta^{AB}+\widebar\psi^A\gamma^{ab}\psi^B\delta_{AB}\right) \,, \\
    T^a &\equiv \hat{\mathcal{D}}V^a - i\widebar\psi_A\gamma^a\psi^A = 0 \,, \\
    F &\equiv dA -\widebar\psi_A\psi_B\epsilon^{AB }-\widebar\psi^A\psi^B\epsilon_{AB } = 0\,, \\
 \label{dgpsi_A}
    \hat{\nabla}\psi_A &\equiv \hat{\mathcal{D}}\psi_A + \frac 1{2\ell} A \psi_B \epsilon_{AC}\delta^{BC} =  + \frac{i}{2\ell}\gamma_a\psi^B V^a\delta_{AB}\,, \\
  \label{dgpsi^A}  \hat{\nabla}\psi^A &\equiv \hat{\mathcal{D}}\psi^A + \frac 1{2\ell} A \psi^B \epsilon^{AC} \delta_{BC} =  + \frac{i}{2\ell}\gamma_a\psi_B V^a \delta^{AB} \,.
\end{align}
\end{subequations}
In the following, we are going to discuss the effects of the deformed background on the supersymmetry properties of  the matter multiplets considered in this paper.

\subsection{Charged matter fields in superspace}
\label{gama}

Let us first consider the immersion in the curved background of a set of hypermultiplets, and then see what does change if we have instead the double-tensor multiplets. 

The hypermultiplets are on-shell multiplets and, as we have seen, in the flat background  they are on-shell equivalent to the double-tensor multiplets. In gauged, matter-coupled  supergravity, when we consider a symplectic covariant gauging \cite{DallAgata:2003sjo,Sommovigo:2004vj,
DAuria:2004yjt,Sommovigo:2005fk,Andrianopoli:2011zj,Trigiante:2016mnt}, dualizing some of the charged scalars of the hypermultiplets into tensors amounts to change the symplectic frame of the electric and magnetic set of gauge fields. This implies in particular that, given an electrically charged scalar in the hypermultiplets, its Hodge-dual tensor is instead electrically charged with respect to the Hodge-dual gauge field, which is then ill-defined. Indeed, the gauge-coupled  tensor undergoes the anti-Higgs mechanism, where the coupling to the gauge field makes the tensor massive, its mass behaving as a magnetic charge.

However, in our case we are not coupling the matter multiplets to $\mathcal{N}=2$ dynamical gauge multiplets, and we consider supergravity as a rigid background, where in particular the graviphoton is an external background field associated with an Abelian gauge connection of $SO(2)\subset SU(2)_R$.

As we are going to see, this introduces a relevant difference between hypermultiplets and double-tensor multiplets, allowing this gauging deformation in the former case, but not in the latter.

\subsubsection{Hypermultiplets in $OSp(2|4)$ rigid superspace}
\label{curvhyp}

Let us consider a set of free hypermultiplets in rigid superspace. As we have seen in Section \ref{hypers}, in the simplest description, with the same notation introduced there, the $n$ hypermultiplets can be expressed in terms of  4$n$ scalars $q^{A\upalpha}$, and of 2$n$ Majorana spinors whose left-handed and right-handed components are $\zeta_\upalpha,\zeta^\upalpha$ respectively. 
In the \emph{flat} rigid superspace background, their field strengths can be defined as
\begin{subequations}
\label{defhypermflat}
    \begin{align}
    \mathcal U^{A\upalpha}&\equiv dq^{A\upalpha}\,,\\
    \nabla \zeta^\upalpha &\equiv d \zeta^\upalpha + \frac 14 \omega^{ab}\gamma_{ab}\zeta^\upalpha\,,\\
    \nabla \zeta_\upalpha &\equiv d \zeta_\upalpha + \frac 14 \omega^{ab}\gamma_{ab}\zeta_\upalpha \,,
    \end{align}
\end{subequations}
and they satisfy on-shell the Bianchi identities
\begin{subequations}
\begin{align}
    d\mathcal U^{A\upalpha}&=0\,,\\
    \nabla^2 \zeta^\upalpha &=0\,,\\
    \nabla^2 \zeta_\upalpha &=0\,,
    \end{align}
\end{subequations}
if their superspace parametrizations, that is their expressions as 1-forms in superspace, are
\begin{subequations}
\label{paramhyperflat}
    \begin{align}
    \mathcal U^{A\upalpha}&= \mathcal U^{A\upalpha}_a V^a + \widebar\psi^A \zeta^\upalpha + \epsilon^{AB}\mathbb{C}^{\upalpha\upbeta} \widebar\psi_B \zeta_\upbeta\,,\\
    \nabla \zeta^\upalpha &= \nabla_a \zeta^\upalpha V^a + i \mathcal U^{A\upalpha}_a \gamma^a\psi_A\,,\\
    \nabla \zeta_\upalpha &= \nabla_a \zeta_\upalpha  V^a + i\epsilon_{AB}\mathbb{C}_{\upalpha\upbeta}  \mathcal U^{B\upbeta}_a \gamma^a\psi^A\,.
    \end{align}
\end{subequations}

In the   $OSp(2|4)$ background, however, the Bianchi identities in superspace get modified and are not anymore satisfied by the above definitions \eqref{defhypermflat} and parametrizations \eqref{paramhyperflat}.
This is due to the fact that, in
the deformed background, the Lorentz-covariant derivative of the gravitino is not zero anymore, being instead
\begin{subequations}
\label{dgpsi}
\begin{align}
\hat{\mathcal{D}}\psi_A=& \frac 1{2\ell}\left(i \gamma_a \psi^B V^a \delta_{AB} - A \psi_B \epsilon_{AC}\delta^{BC}
\right)\,,\\
\hat{\mathcal{D}}\psi^A=& \frac 1{2\ell}\left(i \gamma_a \psi_B V^a \delta^{AB} - A \psi^B \epsilon^{AC}\delta_{BC}
\right)\,,
\end{align}
\end{subequations}
as it follows from eqs. \eqref{dgpsi_A}-\eqref{dgpsi^A}. Let us emphasize that, since the graviphoton gauges a $U(1)\subset SU(2)_R$ in the R-symmetry, then the gauging affects the matter fields charged with respect to  $SU(2)_R$, and in particular the scalars in the hypermultiplets.
Hence, the scalars in the hypermultiplets are charged, with a charge related to the AdS radius $\ell$. More precisely, the modified Lorentz-covariant derivatives \eqref{dgpsi} of the gravitini break the supersymmetry invariance of the theory, which can however be recovered by appropriately deforming the  supercurvatures of the fields.
The modified definitions  of the field strengths are
\begin{subequations}
    \begin{align}
    \hat{\mathcal U}^{A\upalpha}&\equiv \hat \nabla q^{A\upalpha}= dq^{A\upalpha}+ \frac 1{2\ell}A \epsilon^{AB}\delta_{BC}q^{C\upalpha}\,, \label{DefUcurved}\\
    \hat\nabla \zeta^\upalpha &\equiv d \zeta^\upalpha + \frac 14 \gamma_{ab}\hat\omega^{ab}\zeta^\upalpha\,,\\
    \hat\nabla \zeta_\upalpha &\equiv d \zeta_\upalpha + \frac 14 \gamma_{ab}\hat\omega^{ab}\zeta_\upalpha \,.
    \end{align}
\end{subequations}
With  the notations of \cite{Andrianopoli:1996cm}, 
 the  charge deformation corresponds to gauging a translational isometry in the scalar sector, with  scalar  Killing vector  
\begin{align}
k^{A\upalpha}_0\equiv \mathcal{U}^{A\upalpha}_u k^u_0= \frac 1{2\ell}\epsilon^{AB}\delta_{BC}q^{C\upalpha}\,.
\end{align}
The Bianchi identities in superspace are now
\begin{subequations}\label{BIhyperg}
\begin{align}
    \hat\nabla\hat{\mathcal U}^{A\upalpha}&=\frac 1{2\ell}\,dA \epsilon^{AB}\delta_{BC}q^{C\upalpha}\,,\\
    \hat\nabla^2 \zeta^\upalpha &=\frac 14 \gamma_{ab}\hat R^{ab} \zeta^\upalpha\,,\\
    \hat\nabla^2 \zeta_\upalpha &=\frac 14 \gamma_{ab}\hat R^{ab} \zeta_\upalpha\,,
    \end{align}
\end{subequations}
and closure of supersymmetry now requires  their superspace parametrizations to be modified into
\begin{subequations}
    \begin{align}
    \hat{\mathcal U} ^{A\upalpha}&= \hat{\mathcal U}^{A\upalpha}_a V^a + \widebar\psi^A \zeta^\upalpha + \epsilon^{AB}\mathbb{C}^{\upalpha\upbeta} \widebar\psi_B \zeta_\upbeta\,,\\
   \hat \nabla \zeta^\upalpha &= \hat\nabla_a \zeta^\upalpha V^a + i  \hat{\mathcal U}^{A\upalpha}_a \gamma^a\psi_A+ N^\upalpha_A \psi^A\,,\\
    \hat\nabla \zeta_\upalpha &= \hat\nabla_a \zeta_\upalpha  V^a + i\epsilon_{AB}\mathbb{C}_{\upalpha\upbeta}  \hat{\mathcal U}^{B\upbeta}_a \gamma^a\psi^A+ N_\upalpha^A \psi_A\,,
    \end{align}
\end{subequations}
where:
\begin{align}
    \label{shifth}
  N^\upalpha_A=-\frac 1\ell\,\delta_{AB}q^{B\upalpha}\,,\quad N_\upalpha^A=-\frac 1\ell\,\mathbb{C}_{\upalpha\upbeta}\delta_{BC} \epsilon^{AC}q^{B\upbeta}\,.
\end{align}
Moreover, for the Bianchi identities \eqref{BIhyperg} to be satisfied by the parametrizations \eqref{shifth}, the further differential condition has to be satisfied between $ \hat{\mathcal U}^{A\upalpha}_a$ and the spacetime covariant derivative of the spinors:
\begin{align}
    \hat\nabla \hat{\mathcal U}^{A\upalpha}_a = \widebar\psi^B\left(\delta^A_B \hat\nabla_a\zeta^\upalpha + \frac i{2\ell}\gamma_a \zeta_\upbeta\epsilon^{AC}\mathbb{C}^{\upalpha\upbeta}\delta_{BC}
    \right)+
\widebar\psi_B\left(\hat\nabla_a\zeta_\upbeta \epsilon^{AB}\mathbb{C}^{\upalpha\upbeta}+ \frac i{2\ell}\gamma_a \zeta^\upalpha\delta^{AB}
    \right)\,.
\end{align}
We have then shown that, for the hypermultiplets, the embedding in a curved background is possible, and it reproduces  the rigid limit, in the $OSp(2|4)$ background, of the results from gauged supergravity \cite{Andrianopoli:1996vr}.

\subsubsection{Double-tensor multiplets in $OSp(2|4)$ rigid superspace}

For the double-tensor multiplets, instead, the immersion in a curved but rigid supersymmetric background turns out not to be possible.
We can grasp the problem by recalling that, just as we observed for the case of the hypermultiplets,  the deformed background implies  in particular  that the Lorentz-covariant derivative of the gravitino is not zero anymore, being instead given by eqs. \eqref{dgpsi}.
As a consequence, in the presence of the deformed background, the definitions of the super-field strengths get modified.
As we have seen in \eqref{DefUcurved}, for the case of the (undualized) hypermultiplets
compatibility with the curved background implies the following peculiar (\emph{gauging}) deformation on the scalar sector:
\begin{align}
     \hat\nabla q^{\alpha A} = dq^{\alpha A} + \frac 1{2\ell} A\epsilon_{AB}q^{\alpha B}\,,
\end{align}
that is, in terms of the scalar  fields appearing in the double-tensor multiplets, $q^{\alpha A}=\{q^{IA}=(L^I, iM^I)$, $q^{\Ibar A}=(\widebar L^\Ibar, i\widebar M^\Ibar)\}$:
\begin{align}
    \hat E^I &\equiv dL^I + \frac i{2\ell}\,A M^I \,,\label{dhatl}\\
     \hat \nabla M^I&\equiv dM^I + \frac{i}{2\ell}\,A\,L^I \,.
\end{align}
The complex conjugates undergo analogous deformations:
\begin{align}
    \hat {\widebar E}^\Ibar &\equiv d\widebar L^\Ibar -\frac i{2\ell} A \widebar M^\Ibar \,,\\
     \hat {\nabla}\widebar M^\Ibar &\equiv d\widebar M^\Ibar -\frac i{2\ell} A \widebar L^\Ibar\,.
\end{align}
Eq. \eqref{dhatl} implies, in particular, that the gauged scalar vielbein $\hat E^I$ should include  the scalars that have been Hodge-dualized into antisymmetric tensors.

Correspondingly, their Bianchi identities get  modified, into:
\begin{align}
    \hat\nabla \hat E^I=\frac i{2\ell}\,\hat M^I\,dA \,.
\end{align}
It is in fact possible to find a consistent deformation of the scalars parametrizations in superspace, that is
\begin{align}\label{lgau}
    \hat E^I&\equiv E^I + \frac{i}{2\ell}M^I A \,, 
\end{align}
where, to satisfy the Bianchi identities of the scalars $L^I$, the \emph{gauged} field strengths of the (dualized) scalars $M^I$ have to be:
\begin{align}\label{mgau}   \hat \nabla M^I&\equiv dM^I + \frac{i}{2\ell}L^I A \,,
\end{align}
and the parametrizations of the spinors should acquire a fermion shift:
\begin{subequations}
\label{paramzetadef} 
\begin{align}   \hat{\mathcal{D}}\zeta_{\Ibar} &= \hat{\mathcal{D}}_a \zeta_{\Ibar} V^a  +\delta_{I\Ibar}\left(i E^I_a \gamma^a\psi^2 -\frac 32 \,i h^I_a \gamma^a\psi^1\right) + N_\Ibar^A\psi_A\,, \\
    \hat{\mathcal{D}}\zeta^I &= \hat{\mathcal{D}}_a \zeta^I V^a + i E^I_a \gamma^a\psi_1 + \frac 32\,i h^I_a \gamma^a\psi_2 + N^I_A\psi^A\,,
\end{align}
\end{subequations}
where,  as for the hypermultiplets:
\begin{align}
   N^I_A=- \frac 1\ell \left( L^I \delta_{A1} +i M^I\delta_{A2} \right)\,, \quad  N_\Ibar^A= -\delta_{I\Ibar}\frac 1\ell \left( L^I \delta_{A2} -i M^I\delta_{A1}\right) \,.
\end{align}

On the other hand, closure of the 2-form Bianchi identities is problematic. To compensate for the modified covariant derivatives of the gravitino background, it turns out that  the gauged field strengths of the 2-forms $B^I$ should be defined as:
\begin{align}\label{bgau}
    \hat H^I&\equiv H^I + \frac{i}{2\ell}\,A\,\tilde B^I \,.
\end{align}
In the above equation, the tensors $\tilde B^I_{\mu\nu}$, and their field strengths $\tilde H^I_{\mu\nu\rho}$ are new tensor fields, different from the $B^I_{\mu\nu}$, $H^I_{\mu\nu\rho}$.  Their field strengths $\tilde H^I_{\mu\nu\rho}$ are Hodge-dual to the field strengths of the scalars $L^I$, and read (in the flat-space case):
\begin{align}
    \tilde H^I &\equiv d\tilde B^I - 8 M^I \widebar \psi^1 \gamma_a\psi_2 V^a +4 i L^I \left(\widebar\psi_1\gamma_a\psi^1 -\widebar\psi_2\gamma_a\psi^2\right)V^a\,,
    \end{align}
 with the corresponding 
 parametrizations (in flat superspace) being
 \begin{align}\label{tildehpar}
  \tilde H^I&=\tilde H^I_{abc}V^aV^bV^c+2\left(\widebar\psi^1\gamma_{ab}\zeta^I -\widebar\psi_2\gamma_{ab}\zeta_\Ibar \delta^{I\Ibar}\right)V^aV^b\,.
\end{align}  
This, however, requires the further identifications, to be implemented on-shell on the spinors parametrizations:
\begin{align} E^I_a = \frac 32 {\tilde h}^I_a= \frac 14 \epsilon_{abcd}\tilde H^{I | bcd}\,,\quad  h^I_a = -\frac 23 \partial_a M^I \,.\label{duall}
\end{align}
These quantities would appear naturally if we had decided to dualize the $L^I$ instead of the $M^I$ from the very beginning.

All of the modifications above hint at the fact that the coupling with the curved background (which is charged under the graviphoton background field)  implies the ``resurgence" of the Hodge-duals of \emph{all} the fields involved, namely to consider the (complex) enlarged multiplet 
$$\tilde \upphi \equiv \left(L^I,M^I,B^I,\tilde B^I, \zeta^I,\zeta_\Ibar\right)\,,$$
which enjoys matching of d.o.f. only on-shell, where $M^I$ and $B^I$, but also $L^I$ and $\tilde B^I$, are related by Hodge-duality.
However, even taking this into account, the Bianchi identities of the tensor fields do not close the supersymmetry algebra.

We conclude that, differently from the hypermultiplets,   for the double-tensor multiplets it is not possible to find a superspace parametrization allowing to close the supersymmetry algebra in the curved background. There are several ways to understand this result. 
From eqs. \eqref{mgau}, \eqref{bgau} one is to observe that both the fields $M^I$ and $B^I$ should be \emph{electrically} charged with respect to the background gauge field $A$. This is in contrast with the fact that they are  Hodge-dually related (see, e.g., \cite{Trigiante:2016mnt}), as mentioned at the beginning of Section \ref{gama}.
A way out is to assume instead  the $B^I$ to be \emph{magnetically} charged under $A$, that is \emph{electrically} charged under the dual gauge vector $\tilde A$, which is not well-defined, being associated with a magnetic charge (see eq. \eqref{mag}). However, this would imply that in this case the electrically charged fields would be the scalars $L^I$ and  $M^I$, which would bring us to the case of the hypermultiplets, considered in section \ref{curvhyp}.

A group-theoretical way to understand this result is to observe that
the $U(1)$ symmetry gauged by the background field $A$ does not preserve the structure of the double-tensor multiplets, being incompatible with their complex structure.
This is evident referring to the  formulation of the double-tensor multiplets in terms of dualized hypermultiplets (cf. Section \ref{curvhyp}).
Indeed,
the complex structure of the double-tensor multiplets
\begin{subequations}\begin{align}
    (L^I)^
*= \widebar L^\Ibar \quad &\longleftrightarrow \quad (q^{1I})^*= q^{2 \Ibar} \,, \\
(B^I)^
*= \widebar B^\Ibar \quad &\longleftrightarrow \quad (q^{2I})^*= -q^{1 \Ibar}
\end{align}\end{subequations}
is defined by the \emph{diagonal} $U(1)$ factor in the $U(1)_R\times U(n)$ symmetry of the theory, where $U(1)_R\subset SU(2)_R$ is the subset of the R-symmetry kept manifest by the double-tensor multiplet representation,  and acting on the R-symmetry indices $A=1,2$, while $U(n)\subset Sp(2n)$ is associated with the indices $I,\Ibar$.
\\
On the other hand, $A$ is the gauge connection of \emph{another} $U(1)$ group, that is of the $U(1)_R\subset SU(2)_R$,  the subgroup of the R-symmetry preserved by the curved background. In other words,  $A$ does not act on the $U(n)$ factor, making the two symmetries incompatible.

The situation is totally different if the theory is in interaction with a set of dynamical gauge multiplets, and/or at the supergravity level. In both cases, when the gauge fields and their magnetic duals are dynamical fields, it is perfectly possible to construct gauge covariant couplings of the tensors to them. In particular, at the supergravity level this analysis was extensively carried out in, e.g., \cite{DallAgata:2003sjo,Sommovigo:2004vj,DAuria:2004yjt,Sommovigo:2005fk,Andrianopoli:2011zj,Trigiante:2016mnt}.
A general study of double-tensor multiplets coupled with gauge multiplets in curved rigid superspace will be the object of a future investigation.

\section{Conclusions}\label{concl}

Our study of $\mathcal{N}=2$ double-tensor multiplets  explicitly showed that, despite the  matching of bosonic and fermionic degrees of freedom both off- and on-shell, the double-tensor multiplets close the supersymmetry algebra only on-shell.
Moreover we found, unexpectedly, that the scalars $M^I$, Hodge-duals to the antisymmetric tensors $B^I_{\mu\nu}$ in the multiplet, explicitly appear naked in the off-shell definition of the superfields. 
We propose that this may be the origin of the mismatch between power-counting and effective closure off-shell of supersymmetry on these representations.
Indeed, since the Hodge-duality relation between the scalars $M^I$ and the tensors  $B^I_{\mu\nu}$ emerges only on-shell, the   scalars $M^I$ are off-shell independent of the tensors $B^I$, and should then be included in the off-shell counting of degrees of freedom. They always satisfy on-shell    harmonic field equations (see eq. \eqref{327}). 

This is reminiscent of what happens in chiral ($4n+2$)-dimensional supersymmetric models, where the on-shell matching of degrees of freedom requires the ($2n+1$)-form field strengths of the $2n$-indices antisymmetric  tensors 
to be  self-dual. 
In these theories, however, the self-duality constraint is added ``by hand", on the on-shell solutions, since the kinetic term of self-dual ($2n+1$)-forms is automatically vanishing. A dynamical  way to implement the self-duality constraint, from an action principle, was  introduced by A. Sen \cite{Sen:2015nph} (see also \cite{Lambert:2019diy,Hull:2023dgp}), and extended to superspace in \cite{Andrianopoli:2022bzr}. 
This however requires to include off-shell new, unphysical, d.o.f. (in that case, extra antisymmetric tensor fields), which however satisfy on-shell harmonic  equations, and are then fully decoupled from the physical spectrum. It would be interesting to better understand the possible relations between our case and the $(4n+2)$-dimensional chiral models of \cite{Lambert:2019diy,Hull:2023dgp,Andrianopoli:2022bzr}.

The double-tensor multiplets considered here are a particular case belonging to the general family of $\mathcal N=2$ tensor multiplets, which was extensively  studied at the supergravity level in \cite{Louis:1996ya,DallAgata:2003sjo,Sommovigo:2004vj,DAuria:2004yjt,Sommovigo:2005fk,Andrianopoli:2011zj}. In the above mentioned papers, full covariance with respect to the maximal $SU(2)\times Sp(2n)$ holonomy of the quaternionic sector before dualization was kept.
Then, a next step in our investigation will be the extension of our  explicit framework to the supergravity case, 
where we would like to clarify if the extra scalars do play a role also in that case, or if 
this phenomenon, that we observed in the rigid theory, should be relaxed in the more general supergravity context, allowing closure of supersymmetry also in the absence of the dualized scalars.

A further possible extension of our analysis is the coupling of the double-tensor multiplets to a set of gauge multiplets in rigid  superspace. 

The results presented in this paper are intended as a preliminary investigation towards exploring alternative off-shell formulations of these multiplets,  possibly with a finite number of auxiliary fields in an extension of superspace different from the harmonic (or the projective) one.
A promising tool in this direction is the use of the so-called (dual formulation of the) \emph{hidden superalgebras} \cite{DAuria:1982uck} underlying supersymmetric theories in the presence of higher $p$-forms ($p$-indices antisymmetric tensor fields). These superalgebras define an extension of superspace which includes not only  extra even directions but also extra odd directions associated with spin-$3/2$ fields \cite{DAuria:1982uck} (see also \cite{Andrianopoli:2016osu,Andrianopoli:2017itj,Ravera:2018vra,Giotopoulos:2024ovz}). Such extension might allow to circumvent the hypothesis of the no-go theorem on auxiliary fields in $\mathcal{N}$-extended supersymmetry, which indeed contemplates only spin-$1/2$ fields. 
A preliminary result in this direction is provided by the study presented in \cite{Concha:2018ywv}, where  a hidden superalgebra \cite{Ravera:2018vra} underlying $\mathcal{N}=1$, $D=4$ ``flat" supergravity (meaning in the absence of any internal scale in the Lagrangian), was found. {In that case, the bosonic and fermionic 1-forms dual to the generators of the hidden superalgebra} play the specific role of ``topological auxiliary fields": Their inclusion, by means of boundary terms in the theory, restores supersymmetry when a nontrivial spacetime boundary is present. Moreover, as later observed in \cite{Andrianopoli:2021rdk}, they serve as auxiliary fields for the (off-shell) bulk theory. Specifically, the field equations for these fields enforce the Bianchi identities of the Lorentz spin connection and gravitino 1-form fields.
In this context, the rheonomic approach consistently proves to be a valuable asset in the construction of both rigid supersymmetric theories and supergravities.

Another approach for circumventing the no-go theorems, at the supergravity level, involves deriving $\mathcal{N}=2$ supergravity coupled with a single hypermultiplet using the \emph{double copy} prescription described in \cite{Diaz-Jaramillo:2021wtl,Bonezzi:2022yuh,Bonezzi:2022bse,Bonezzi:2023lkx} for pure $\mathcal{N}=1$ super Yang-Mills, framed within the context of \emph{homotopy algebras}. This prescription aims to provide an off-shell and gauge-independent theory valid to all orders of interaction (currently demonstrated up to the third order) in generic spacetime dimensions. Although at present this scheme has only been studied for bosonic theories, a first example of its application to a \emph{supersymmetric double copy} will be soon available in the upcoming paper \cite{Bonezzi:2024Upcoming} by one of the authors. Some remaining challenges in this approach would be to extend the technique to superspace and to incorporate in the theory an arbitrary number of hypermultiplets, tensor multiplets or, in general, matter multiplets not belonging to the adjoint representation. It would also be interesting to see whether the resulting set of auxiliary fields in superspace corresponds to the one proposed by the hidden superalgebras formulation described above.

\section*{Acknowledgments}

We acknowledge interesting discussions with Roberto Bonezzi, Riccardo D'Auria,  Olaf Hohm, Mario Trigiante and Rikard von Unge. The work of G.C. is funded by the Deutsche Forschungsgemeinschaft (DFG, German Research Foundation), Projektnummer 417533893/GRK2575 “Rethinking Quantum Field Theory”. G.C. would also like to thank INFN Torino and Polytechnic of Turin for hospitality.

\appendix

\section{Conventions, notation and useful formulas}
\label{conventions}

In the following, we are going to report the set of conventions that have been used in our work. In particular, we will present our notations for $2$-component and $4$-component Majorana spinors as well as for Dirac spinors.
Furthermore, we write down some Fierz identities, that were used to find our results.

\subsection{Conventions on spinors}

We denote with
$\Psi$, $\Psi_A$ ($A=1,2\in U(2)$) the (4-components) Majorana spinor gravitini in $\mathcal{N}=1$ and $\mathcal{N}=2$, respectively.
The Majorana spinor condition, on a general spinor $\lambda$, is
\begin{equation}
    \widebar \lambda =\lambda^t \,C \,,
\end{equation}
where the adjoint spinor $\widebar\lambda$ is defined as
\begin{equation}
    \widebar \lambda \equiv \lambda^\dagger \,\gamma^0\,.
\end{equation}
The Weyl projections on the left and right chiral components of the spinors are given in terms of the projectors
\begin{equation}
    \mathbb{P}_\pm \equiv \tfrac{1}{2}\left(\mathbb{I} \pm \gamma_5\right)\,:
\end{equation}
\begin{equation}\begin{split}
    \psi_\bullet \equiv \mathbb{P}_+ \Psi \quad  &\to  \quad \mathcal{N}=1 \quad \text{Chiral Majorana spinor} \,, \\
    \psi^\bullet \equiv \mathbb{P}_- \Psi \quad  &\to  \quad \mathcal{N}=1 \quad \text{Anti-chiral Majorana spinor}\,,\\
    \psi_A \equiv  \mathbb{P}_+\Psi_A \quad  &\to  \quad \mathcal{N}=2 \quad \text{Chiral Majorana spinors} \,,\\
    \psi^A \equiv \mathbb{P}_-\Psi_A  \quad  &\to  \quad \mathcal{N}=2 \quad \text{Anti-chiral Majorana spinors}\,.
\end{split}
\end{equation}
We define the $\mathcal{N}=2$  Dirac gravitino 1-form and its complex conjugate:
\begin{align}
    \uppsi \equiv \Psi_1 + i\Psi_2\,; \quad \uppsi^* \equiv \Psi_1 - i\Psi_2\,.
\end{align}
Their adjoint spinor-1-forms are 
\begin{align}
    \widebar\uppsi =  \widebar\Psi_1 - i\widebar\Psi_2\,; \quad \widebar\uppsi^* =  \widebar\Psi_1 + i\widebar\Psi_2 \,. 
\end{align}
They are related to the  Majorana gravitini 1-forms by
\begin{equation}\begin{split}
    \Psi_1 &= \tfrac{1}{2}(\uppsi+\uppsi^*)\,; \quad\;\;\, \widebar\Psi_1 = \tfrac{1}{2}(\widebar\uppsi+\widebar\uppsi^*)\,; \\
    \Psi_2 &= -\tfrac{i}{2}(\uppsi-\uppsi^*)\,; \quad \widebar\Psi_2 = \tfrac{i}{2}(\widebar\uppsi-\widebar\uppsi^*)\,.
\end{split}\end{equation}
Finally, it is convenient to introduce the Dirac spinor 1-form combinations
\begin{equation}
    \begin{split}
        \uppsi_\pm &\equiv \frac 12 \left(\uppsi \pm \gamma_5\uppsi^*\right)= \mathbb{P}_\pm \Psi_1 + i \mathbb{P}_\mp \Psi_2\,:\quad \left\{ \begin{matrix}
           \uppsi_+ = \psi_1 + i \psi^2\\
           \uppsi_- = \psi^1 + i \psi_2
        \end{matrix} \,,
        \right.
    \end{split}
\end{equation}
and their adjoint,
\begin{equation}
    \begin{split}
        \widebar\uppsi_\pm &\equiv \frac 12 \left(\widebar\uppsi \pm \widebar\uppsi^*\gamma_5\right)= \widebar\Psi_1\mathbb{P}_\pm  - i \widebar\Psi_2\mathbb{P}_\mp\,:\quad \left\{ \begin{matrix}
           \widebar\uppsi_+ = \widebar\psi_1 - i \widebar\psi^2\\
           \widebar\uppsi_- = \widebar\psi^1 - i \widebar\psi_2
        \end{matrix} \,.
        \right.
    \end{split}
\end{equation}
They satisfy
\begin{align}
   \widebar{\left(\uppsi_\pm\right)} \equiv\left(\uppsi_\pm\right)^\dagger \gamma_0= \widebar\uppsi_\mp \,.
\end{align}

\subsection{Properties of gamma-matrices}

The four-dimensional Dirac matrices are defined as
\begin{equation}
    \gamma^a \equiv \begin{pmatrix} 0 & \sigma^a \\ \widebar\sigma^a & 0 \end{pmatrix}\,, \quad \gamma_5 \equiv -\frac{i}{4!}\epsilon_{abcd}\gamma^a\gamma^b\gamma^c\gamma^d\,, \quad \gamma^{ab} \equiv \gamma^{[a} \gamma^{b]}=\begin{pmatrix}  \sigma^{[a}\widebar\sigma^{b]}&0 \\  0 &\widebar\sigma^{[a}\sigma^{b]}\end{pmatrix} \,,
\end{equation}
with $\sigma^a \equiv (\mathbb{I}, \sigma^x)$, $\widebar\sigma^a \equiv (\mathbb{I}, -\sigma^x)$,  where  $\sigma^x$ are the Pauli matrices and $x=1,2,3$. The Dirac gamma matrices fulfill
\begin{subequations}\begin{align}
    \gamma_0^\dagger &= \gamma_0 \,,\\
    \gamma_0\gamma^a\gamma_0 &= (\gamma^a)^\dagger \,,\\
    \gamma_5^\dagger &= \gamma_5 \,,\\
    \gamma_5^* &= \gamma_5 \,, \\
    (\gamma_5)^2 &= \mathbbm{1} \,,\\
    \{\gamma^a,\gamma^b\} &= 2\eta^{ab} \,,\\
    [\gamma^a,\gamma^b] &= 2\gamma^{ab} \,,\\
    \gamma^a\gamma^b &= \eta^{ab} + \gamma^{ab}\,.
\end{align}\end{subequations}
Moreover, they respect the following duality properties:
\begin{equation}\begin{split}
    \gamma_a = -\frac{i}{3!}\epsilon_{abcd}\gamma_5\gamma^{bcd}\,&, \quad \gamma_{ab} = -\frac{i}{2}\epsilon_{abcd}\gamma_5\gamma^{cd}\,, \\
    \gamma_{abc} = i\epsilon_{abcd}\gamma_5\gamma^d\,&, \quad \gamma_{abcd} = i\epsilon_{abcd}\gamma_5
\end{split}\end{equation}
and show some remarkable products, such as:
\begin{subequations}\begin{align}
    \gamma_a\gamma^a &= 4 \,,\\
    \gamma_{ab}\gamma^{ab} &= -12 \,,\\
    \gamma_{ab}\gamma^c &= 2\gamma_{[a}\delta_{b]}^c + i\epsilon_{ab}{}^{cd}\gamma_5\gamma_d \,,\\
    \gamma_{ab}\gamma^{cd} &= \gamma_{ab}{}^{cd} - 4\delta^{[c}_{[a}\gamma_{b]}{}^{d]} - 2\delta_{ab}^{cd} \,,\\
    \gamma_b\gamma^{a_1\dots a_k}\gamma^b &= 2(2-k)(-1)^k\gamma^{a_1\dots a_k} \,,\\
    \gamma_a\gamma_m\gamma^a &= -2\gamma_m \,, \\
    \gamma_{ab}\gamma^a &= -3\gamma_b \,,\\
    \gamma_a\gamma_{mn}\gamma^a &= 0 \,,\\
    \gamma_{ab}\gamma_m\gamma^{ab} &= 0 \,,\\
    \gamma_{ab}\gamma_{mn}\gamma^{ab} &= 4\gamma_{mn}\,.
\end{align}\end{subequations}
The charge conjugation matrix $C= \begin{pmatrix}
    \epsilon^{\alpha\beta}&0\\
    0&\epsilon_{\dot\alpha\dot\beta}
\end{pmatrix}$ has the following properties:
\begin{equation}\begin{split}
    C^2=-1\,&, \quad C^T=-C\,, \quad (C\gamma_a)^T=C\gamma_a\,, \quad (C\gamma_5)^T=-C\gamma_5\,, \\
    &(C\gamma_5\gamma_a)^T= -C\gamma_5\gamma_a\,, \quad (C\gamma_{ab})^T=C\gamma_{ab} \,.
\end{split}\label{Eq:ChargeConjMatrix}
\end{equation}

\subsection{Useful Fierz identities}\label{Fierzapp}
In the following we provide a list of useful Fierz identities for spin-$\frac12$ and spin-$\frac32$ fermions in an $\mathcal{N}=2$ supersymmetric theory.

\subsubsection*{Fierz identities for two Majorana fermions}

For 1-form fermions (gravitini):
\begin{subequations}\begin{align}
    \psi_A\widebar\psi_B &= \frac{1}{2}\widebar\psi_B\psi_A - \frac{1}{8}\gamma_{ab}\widebar\psi_B\gamma^{ab}\psi_A \,,\\
    \psi^A\widebar\psi^B &= \frac{1}{2}\widebar\psi^B\psi^A - \frac{1}{8}\gamma_{ab}\widebar\psi^B\gamma^{ab}\psi^A \,,\\
    \psi_A\widebar\psi^B &= \frac{1}{2}\gamma_a\widebar\psi^B\gamma^a\psi_A \,,\\
    \psi^A\widebar\psi_B &= \frac{1}{2}\gamma_a\widebar\psi_B\gamma^a\psi^A \,.
\end{align}\end{subequations}
For 0-form fermions (left-handed $\lambda_\bullet, \xi_\bullet$, or right-handed $\lambda^\bullet, \xi^\bullet$):
\begin{subequations}\begin{align}
    \lambda_\bullet\widebar\xi_\bullet &= - \frac{1}{2}\widebar\xi_\bullet\lambda_\bullet + \frac{1}{8}\gamma_{ab}\widebar\xi_\bullet\gamma^{ab}\lambda_\bullet \,,\\
\lambda^\bullet\widebar\xi^\bullet &= - \frac{1}{2}\widebar\xi^\bullet\lambda^\bullet + \frac{1}{8}\gamma_{ab}\widebar\xi^\bullet\gamma^{ab}\lambda^\bullet \,,\\    \lambda_\bullet\widebar\xi^\bullet &= - \frac{1}{2}\gamma_a\widebar\xi^\bullet\gamma^a\lambda_\bullet \,.
\end{align}\end{subequations}

\subsubsection*{Fierz identities for three Majorana gravitini}
For three 1-form gravitini we find
\begin{subequations}
\begin{align}
&\gamma_a\psi^C\widebar\psi_A\gamma^a\psi^B {= -\gamma_a\psi^B\widebar\psi_A\gamma^a\psi^C= 2 \psi_A\widebar\psi^B\psi^C\neq 0 } \,, \\
&\gamma_{ab}\psi_{(A}\widebar\psi_B\gamma^{ab}\psi_{C)} = 0 \,, \\
&{\gamma_{ab}\psi_{[A}\widebar\psi_{B]}\gamma^{ab}\psi_{C} =6\psi_C\widebar\psi_A\psi_B } \,, \\  &\gamma_{ab}\psi_A\widebar\psi^B\gamma^{ab}\psi^C = 0  \,,
\end{align}
\end{subequations}
which are particular cases of
\begin{subequations}
\begin{align}
\gamma_{ab}\psi_A\widebar\psi^B\gamma^{ab}\psi^C &= 0 \,,\\
\gamma_a\psi^{(A}\widebar\psi^{B)}\gamma^a\psi_C &= 0 \,,\\
\gamma_a\psi_{(A}\widebar\psi_B\gamma^a\psi_{C)} &= 0 \,,\\
\gamma_{ab}\psi_{(A}\widebar\psi_B\gamma^{ab}\psi_{C)} &= 0 \,.
\end{align}
\end{subequations}



\section{$\mathcal{N}=2$ super-Poincar\'e algebra and its dual formulation}\label{superalgebraapp}
Here we show the relation between the (dual) formulation of the superalgebra, expressed in terms of \emph{Maurer-Cartan equations}, and the standard formulation of the same superalgebra, usually written down by means of \emph{graded commutators}.

The  $\mathcal{N}=2$ Poincar\'e superalgebra underlying the rigid supersymmetric background is given in terms of the following non-vanishing (anti-)commutators:
\begin{subequations}
\label{superalgapp}
\begin{align}
    [J_{ab},p_c] &= -2\eta_{c[a}p_{b]} \,,\\
    [J_{ab},J^{cd}] &= -4\delta^{[c}_{[a}J_{b]}^{\;\;d]} \,,\\
    {[J_{ab},Q^{ \alpha A}]} &= \frac 12(\gamma_{ab})^{\alpha}{}_{\beta}Q^{\beta A} \,,\\{[J_{ab},Q_{\dot\alpha A}]} &= \frac 12(\gamma_{ab})_{\dot\alpha}{}^{\dot\beta}Q_{\dot\beta A} \,,\\
    \{ Q^{\alpha A},Q_{\dot\beta B}\} &=  -i \delta^A_B\,(C \gamma^a)^\alpha{}_{\dot\beta}\, p_a\,, 
    \\
   {\{Q^{\alpha A},Q^{\beta B}\}} &  =   C^{\alpha\beta}\epsilon^{AB}Z 
    \,,,\\
    {\{Q_{\dot\alpha A},Q_{\dot\beta B}\}} & =  C_{\dot\alpha\dot\beta}\epsilon_{AB}   \bar Z \,,
    \\
    [T^x,T^y] &= \epsilon^{xyz} T^z \,,
\end{align}
\end{subequations}
where $p_a$ and $J_{ab}$ are the translation and the rotation generators respectively,  $Q^{A\alpha}, Q_{A\dot\alpha}$ are the left- and right-handed chirality projections of the supersymmetry generators, satisfying
\begin{equation}
    \gamma_5 \begin{pmatrix}
Q^{A }\cr Q_{A } \end{pmatrix}   
=\begin{pmatrix}Q^{A }\cr -Q_{A }    
\end{pmatrix}\,.
\end{equation} 
The superalgebra also includes the $SU(2)\times U(1)$ R-symmetry generators $T^x_{AB}, T^0_{AB}$, under which the odd generators  $Q_A,Q^A$ of the super-Poincar\'e algebra are not-charged. A  linear combination  of them, $Z(T^x,T^0)$, behaves as a complex central charge of the supersymmetry algebra. With a small abuse of notation, we  denoted:
\begin{align}   (C\gamma^a)^{\alpha}{}_{\dot\beta}\equiv \epsilon^{\alpha\beta}(\sigma^a) _{\beta\dot\beta}
\,.
\end{align}

It is possible to express the superalgebra in a dual form, in terms of the Maurer-Cartan 1-forms $$\sigma^{\mathcal A}\equiv (V^a, \omega^{ab}, A, \tilde A,   \psi_{A\alpha}, \psi^A_\alpha),
$$
where $V^a$ and  $\omega^{ab}$ are the bosonic vielbein 1-form and  Lorentz spin-connection respectively, $A$ an Abelian gauge connection (the graviphoton, which in the rigid background is a pure gauge), $\tilde A$ its Hodge dual magnetic field, and  $\psi_{A }, \psi^{A }$ are the left- and right-handed   chirality projections of the gravitini 1-forms:
\begin{equation}
    \gamma_5 \begin{pmatrix}
\psi_{A }\cr \psi^{A } \end{pmatrix}   
=\begin{pmatrix}\psi_{A }\cr -\psi^{A }    
\end{pmatrix}\,.
\end{equation}
The set of 1-forms $\sigma^{\mathcal A}$ are defined as the Poincar\'e dual of the superalgebra generators
$$f_{\mathcal A}\equiv (p_a, J_{ab}, Z,  Q^{A\alpha}, Q_A^\alpha) $$
satisfying $\sigma^{\mathcal A}(f_{\mathcal B})=\delta^{\mathcal A}_{\mathcal B}$, that is to say
\begin{equation}\begin{matrix}
    V^a(p_b) = \delta^a_b\,,
    & \omega^{ab}(J_{cd}) = 2\delta^{ab}_{cd}   \,,&A(Re(Z))=1\,, &\tilde A (Im (Z))=1\,,\\
    & \psi_{\alpha A}(Q^{\beta B}) = \delta_\alpha^\beta \delta^B_A\,,
    & \psi^{\dot\alpha A}(Q_{\dot\beta B}) = \delta^{\dot\alpha}_{\dot\beta}\delta^A_B  \,,
    &
\end{matrix}\end{equation}
and
\begin{subequations}\begin{align}
    d\sigma^{\mathcal A}(f_{\mathcal B},f_{\mathcal C}) &= -\frac12 \sigma^{\mathcal A}([f_{\mathcal B},f_{\mathcal C}\})\,,\\
    \sigma^{\mathcal A} \!\wedge \sigma^{\mathcal B} \left(f_{\mathcal C},f_{\mathcal D}\right)&=\frac 12 \left(\sigma^{\mathcal A}(f_{\mathcal C})\sigma^{\mathcal B}(f_D)- \sigma^{\mathcal B}(f_{\mathcal C})\sigma^{\mathcal A}(f_\mathcal{D})
    \right)\,,
\end{align}\end{subequations}
where $[\cdot,\cdot\}$ indicates the graded commutator.

Using the above definitions, it is straightforward to  derive the Maurer-Cartan equations  expressing  the superalgebra \eqref{superalgapp} in its dual form. They are:
\begin{subequations}\label{backapp}\begin{align}
    &d\omega^{ab} + \omega^a{}_c \wedge \omega^{cb} = 0 \,, \\
    &dV^a + \omega^{ab} V_b  - i\widebar\psi_A  \gamma^a\psi^A= 0 \,,\\
    &d\psi_A + \frac14\gamma_{ab}\omega^{ab} \psi_A = 0 \,,\\
    &d\psi^A + \frac14\gamma_{ab}\omega^{ab}\psi^A = 0 \,, \\
    &d  A -  \widebar\psi_A \psi_B\epsilon^{AB} -  \widebar\psi^A  \psi^B\epsilon_{AB} = 0 \,, \\
    &d \tilde A -  \widebar\psi_A \psi_B\epsilon^{AB} +  \widebar\psi^A  \psi^B\epsilon_{AB} = 0 
    \,.
\end{align}\end{subequations}
Eqs. \eqref{backapp} are the equations defining our flat, rigid supersymmetric background.

\section{Comparison with $\mathcal{N}=1$ linear multiplets}\label{linN1comparison}

In the $\mathcal{N}=1$ supersymmetry case, it is possible to introduce the so-called \emph{linear multiplets}, which are real multiplets realizing an off-shell representation of $\mathcal{N}=1$ supersymmetry. A general superspace description of free linear multiplets in rigid, flat $\mathcal N=1$ superspace was given in \cite{Bertolini:1994cb}.
To clarify similarities and differences of the double-tensor multiplets with respect to the case of $\mathcal{N}=1$ linear multiplets, let us compare explicitly the expressions for the super-field strengths and Lagrangian in the two models.

\subsection{$\mathcal{N}=1$ linear multiplets in rigid, flat superspace}

The $\mathcal{N}=1$ linear multiplets are real multiplets, each composed by a real scalar $X^I$, a real 2-form $\mathcal B^I$ and a Majorana spinor $\lambda^I$ ($I=1,\dots,m$):
\begin{align}
    \Phi_{\mathcal{N}=1}=\left( X^I, \mathcal B^I,\lambda^I \right)\,.
\end{align}
These multiplets realize an off-shell representation of $\mathcal{N}=1$ supersymmetry. In \cite{Bertolini:1994cb}, the Lagrangian and supersymmetry transformation laws of a set of free linear multiplets in a flat supersymmetric background was constructed, using a general set of coordinates $z^i$ ($i=1,\ldots ,m$ being anholonomic indices on the scalar sigma model) for the scalar sector, and thus expressing it as a non-linear $\sigma$-model. Here, we choose to describe the scalar sector as a linear  $\sigma$-model, picking as coordinates the scalar sections themselves: $z^i\to X^I(z)=X^I$.

The background field content of the model is the following (with the gravitino  $\Psi$ being a Majorana spinor):
\begin{subequations}
\begin{align}
    R^{ab} &\equiv d\omega^{ab} + \omega^a{}_c\wedge\omega^{cb} = 0 \,, \\
    \mathcal{D}V^a &\equiv dV^a + \omega^a{}_b  V^b = \frac{i}{2}\widebar\Psi \gamma^a\Psi \,, \\
    \mathcal{D}\Psi &\equiv d\Psi + \frac{1}{4}\gamma_{ab}\omega^{ab} \Psi = 0 \,.
\end{align}
\end{subequations}
The definitions of the super-field strengths read (in terms of 4-components Majorana spinors  not decomposed in their chiral projections)
\begin{subequations}\begin{align}
    \mathcal{E}^I &\equiv d X^I \,, \\
    \mathcal{H}^I &\equiv d\mathcal{B}^I - i X^I \widebar{\Psi} \gamma_a\Psi V^a \,, \\
    \mathcal{D}\lambda^I &\equiv d\lambda^I + \frac{1}{4} \omega^{ab} \gamma_{ab} \lambda^I \,.
\end{align}\end{subequations}
The corresponding Bianchi identities are
\begin{subequations}\label{BIlm}
\begin{align}
    d\mathcal{E}^I &= 0 \,,\\
    d\mathcal{H}^I &= -i d X^I \widebar\Psi \gamma_a\Psi V^a \,,\\
    \mathcal{D}^2 \lambda^I &= 0\,.
\end{align}
\end{subequations}

\subsubsection*{Off-shell closure in superspace}
One finds that the Bianchi identities \eqref{BIlm} are satisfied by the following rheonomic parametrization of the superfields in superspace:
\begin{subequations}\label{parlm}
\begin{align}
    \mathcal{E}^I &= \mathcal{E}^I_aV^a + \widebar\Psi\lambda^I \,, \\
    \mathcal{H}^I &= \mathcal{H}^I_{abc}V^aV^bV^c + \widebar\lambda^I\gamma_{ab}\Psi V^aV^b \,, \\
    \nabla\lambda^I &= \nabla_a\lambda^IV^a + \left(\frac{i}{2}\mathcal{E}^I_a +  \frac{3}{2} \gamma_5 \mathsf{h}^I_a \right)\gamma^a\Psi\,,
\end{align}
\end{subequations}
where
\begin{align}\label{hHlm}
    \mathsf{h}^I_a = \frac 16 \epsilon_{abcd} \mathcal{H}^{I|bcd} \,.
\end{align}
It is important to emphasize that, for this kind of supermultiplets, the Bianchi identities \eqref{BIlm} are \emph{identically} satisfied by the superfield parametrizations \eqref{parlm},  relying on the $\mathcal{N}=1$ Fierz identities, \emph{without use of the field equations}.

It is instructive to see explicitly how this works, in the complementary spacetime approach, that is to show that in this case, differently from the $\mathcal{N}=2$ double-tensor multiplets, the supersymmetry algebra closes off-shell on the fields.

\subsubsection*{Off-shell closure 
on spacetime 
}

The supersymmetry transformation laws are
\begin{subequations}
    \label{lmsusylaws}
    \begin{align}
        \delta_\epsilon X^I =& \widebar\epsilon\lambda^I\,,\\
    \delta_\epsilon \mathcal{B}^I_{\mu\nu} =& -\widebar\epsilon\gamma_{\mu\nu}\lambda^I\,,\\     \delta_\epsilon \lambda^I = &\left(\frac i2 \partial_\mu X^I+ \frac 32 \mathsf{h}^I_\mu\,\gamma_5
    \right) \gamma^\mu \epsilon\,.
    \end{align}
\end{subequations}
We then find, using the properties of the $\gamma$-matrices given in \eqref{Eq:ChargeConjMatrix}:
\begin{subequations}
    \label{lmclosure}
    \begin{align}
\left[\delta_{\epsilon},\delta_{\sigma}\right]  X^I =& \frac i2 \partial_\mu X^I \left( \widebar\sigma\gamma^\mu \epsilon   - \widebar\epsilon\gamma^\mu \sigma \right)
    +\frac32 \mathsf{h}_\mu ^I\left(\cancel{ \widebar\sigma\gamma^\mu \,\gamma_5\epsilon }-\cancel{  \widebar\epsilon\gamma^\mu \,\gamma_5\sigma }\right) \nonumber \\
    =& -i\partial_\mu X^I\,\widebar\epsilon\gamma^\mu \sigma \,,\\  \left[\delta_{\epsilon},\delta_{\sigma}\right]  \mathcal{B}^I_{\mu\nu} = &
    \frac i2 \partial_\rho X^I\left(-\widebar\sigma\gamma_{\mu\nu}\gamma^\rho\epsilon+\widebar\epsilon\gamma_{\mu\nu}\gamma^\rho\sigma\right)
    -\frac 32 \mathsf{h}^I_\rho\left(-\widebar\sigma\gamma_{\mu\nu}\gamma_5\gamma^\rho\epsilon+\widebar\epsilon\gamma_{\mu\nu}\gamma_5\gamma^\rho\sigma\right)\nonumber \\
    =&- 3i\,\partial_{[\rho}\mathcal{B}_{\mu\nu]}^I \widebar\epsilon\gamma^\rho  \sigma  -i\partial_{[\mu}X^I \widebar\epsilon\gamma_{\nu ]}  \sigma \nonumber \\
    =& -i\partial_\rho \mathcal{B}^I_{\mu\nu}\,\widebar\epsilon\gamma^\rho \sigma + \partial_{[\mu}\Xi^I_ {\nu ]} \,,\\
\left[\delta_{\epsilon},\delta_{\sigma}\right] \lambda^I = &-\frac i2 \,\gamma^\mu\left(\sigma\widebar\epsilon -\epsilon\widebar\sigma 
    \right)\partial_\mu\lambda^I+ \frac 14 \,\gamma_5\gamma^\mu\left(\sigma\widebar\epsilon -\epsilon\widebar\sigma \right) \epsilon_{\mu\nu\rho\sigma}\, \gamma^{\rho\sigma}\partial^\nu\lambda^I \nonumber \\
    =&-\frac i2 \,\gamma^\mu\left(\frac 12 \gamma_\rho \widebar\epsilon\gamma^\rho \sigma -\frac 14 \gamma_{\rho\sigma} \widebar\epsilon\gamma^{\rho\sigma}\sigma\right)\partial_\mu\lambda^I+\nonumber\\
    &+ \frac i2 \,\gamma_5\gamma^\mu\left(\frac 12 \gamma_\rho \widebar\epsilon\gamma^\rho \sigma -\frac 14 \gamma_{\rho\sigma} \widebar\epsilon\gamma^{\rho\sigma}\sigma\right)\,\gamma_{\mu\nu}\gamma_5\partial^\nu\lambda^I \nonumber \\  
    =&- \frac i4 \widebar\epsilon\gamma^\rho \sigma\left[\left(-\cancel{\gamma_\rho \slashed{\partial} \lambda^I}+ 2 \partial_\rho \lambda^I\right) + \left(\cancel{\gamma_\rho \slashed{\partial} \lambda^I}+ 2 \partial_\rho \lambda^I\right)\right] + \nonumber \\
    &+\frac i8 \widebar\epsilon\gamma^{\rho\sigma} \sigma\left[\cancel{\gamma^\nu\gamma_{\rho\sigma}}+ \cancel{\gamma_\mu\gamma_{\rho\sigma}\gamma^{\mu\nu}}\right]\partial_\nu\lambda^I\nonumber\\
    =& -i  \widebar\epsilon\gamma^\rho \sigma\partial_\rho\lambda^I \,,\label{delta2lambda} \end{align}
\end{subequations}
where 
\begin{align}
    \Xi^I_\mu \equiv -2i\,\mathcal{B}_{\mu\nu}^I \widebar\epsilon\gamma^\nu\sigma - iX^I \widebar\epsilon\gamma_{\mu}  \sigma \,.
\end{align}
We remark that, for the cancellation of the terms in $\slashed{\partial} \lambda^I$ (which would spoil the closure off-shell) in \eqref{delta2lambda}, it is crucial that the contributions in
$\widebar\epsilon\gamma^\rho \sigma$ from the first term (coming from $\delta X^I$), cancel against the second ones from the third term (coming from $\delta h_\mu^I$), where we used \eqref{hHlm} (and the same for the terms in $\widebar\epsilon\gamma^{\rho\sigma} \sigma$). 

On the contrary, in the case of the Wess-Zumino multiplets, which can be obtained from the linear multiplets by Hodge-dualizing the tensors, that is by replacing $h^I_\mu \to \partial_\mu Y^I$, $Y^I$ being pseudoscalars, the contributions from $\delta \partial_\mu Y^I$ have the same structure as the ones from $\delta \partial_\mu X^I$, so that the cancellation does not take place.

\subsubsection*{The $\mathcal{N}=1$ linear multiplets superspace Lagrangian}

The rheonomic $\mathcal{N}=1$ linear multiplets Lagrangian in superspace is
\begin{align}
\mathcal{L}^{\mathcal{N}=1}_{\text{superspace}}&=a_1 \delta_{IJ}\tilde{\mathcal{E}}^{I|a}\left(\mathcal{E}^J-\widebar\Psi\lambda^J\right)V^bV^cV^d\epsilon_{abcd} -\frac 18 a_1 \delta_{IJ}\tilde{\mathcal{E}}^{I|\ell}\tilde{\mathcal{E}}^{J }_\ell V^aV^bV^cV^d\epsilon_{abcd}\nonumber\\
&+a_2 \tilde{\mathsf{h}}^I_c\left(\mathcal{H}_I-\delta_{IJ}\widebar\lambda^J\gamma_{ab}\Psi V^a V^b\right)V^c -\frac 18 a_2\tilde{\mathsf{h}}^{I|\ell}\tilde{\mathsf{h}}^{J }_\ell \delta_{IJ}V^aV^bV^cV^d\epsilon_{abcd}\nonumber\\
&+a_3 \delta_{IJ}\widebar\lambda^I\gamma_a\nabla \lambda^J\,V^bV^cV^d\epsilon_{abcd} \nonumber \\
&+b_1 \delta_{IJ}\mathcal{E}^I \widebar\lambda^J\gamma_{ab}\gamma_5 \Psi V^aV^b + b_2 H_I\widebar\lambda^I\gamma_5\Psi \nonumber\\
&+\delta_{IJ}\left(c_{1}\widebar\lambda^I\lambda^J\widebar\Psi \gamma_{ab}\gamma_5\Psi+c_{2}\widebar\lambda^I\gamma_5\lambda^J \widebar\Psi \gamma_{ab}\Psi
\right)V^aV^b \,,
\end{align}
with coefficients
\begin{align}
    a_2= -9 a_1\,, \quad a_3= - i a_1\,, \quad b_1=3 i a_1\,, \quad b_2= -3 i a_1\,, \quad c_1=c_2=\frac 34 i a_1 \,.
\end{align}

\subsection{$\mathcal{N}=2$ in Dirac-spinor formalism}

To clarify similarities and differences of the double-tensor multiplets with $\mathcal{N}=1$ linear multiplets, we find it useful to reformulate the model in terms of Dirac spinors.  

As far as the two (Majorana spinors) gravitini $\Psi_A$ are concerned, they can be expressed in terms of a single (Dirac spinor) gravitino $\uppsi$ and of its complex conjugate $\uppsi^*$, defined as (see also Appendix \ref{conventions} for details):
\begin{align}
    \uppsi\equiv \Psi_1 + i \Psi_2\,,\quad \uppsi^*\equiv \Psi_1 - i \Psi_2\,,
\end{align}
their adjoint spinors being
\begin{align}
    \widebar\uppsi\equiv \uppsi^\dagger \gamma_0= \widebar \Psi_1 -i \widebar \Psi_2\,\quad \widebar\uppsi^*\equiv \uppsi^t \gamma_0= \widebar \Psi_1 +i \widebar \Psi_2\,.
\end{align}
In terms of them, the background  is given by
\begin{subequations}\begin{align}
    \mathcal{R}^{ab} &\equiv d\omega^{ab} + \omega^a{}_c\wedge\omega^{cb} = 0 \,, \\
    T^a &\equiv   dV^a + \omega^a{}_b  V^b - \frac{i}{2}\widebar\uppsi \gamma^a\uppsi =0 \,, \\
    F&\equiv dA -  \widebar\uppsi\uppsi =0\,, \\
    \tilde F&\equiv d\tilde A -  \widebar\uppsi\gamma_5\uppsi =0  \,, \\
    \mathcal{D}\uppsi &\equiv d\uppsi + \frac{1}{4}\gamma_{ab}\omega^{ab} \uppsi = 0 \,,
\end{align}\end{subequations}
while the parametrizations of the hypermultiplet scalars, eqs. \eqref{hypsca}, read
\begin{subequations}\label{hypscapm}
\begin{align}
    E^I\equiv\;\mathcal{U}^{1 I} &= \mathcal{U}^{1 I}_aV^a+ \left( \widebar\uppsi -\widebar\uppsi^* \gamma_5\right)\chi^I = \mathcal{U}^{1 I}_aV^a +   \widebar \chi^I \left( \uppsi - \gamma_5 \uppsi^* \right) 
    \label{e}\\
    &=\mathcal{U}^{1 I}_aV^a +\widebar\psi^1 \zeta^I +\widebar\psi_2\zeta_\Ibar \delta^{I\Ibar}\,,
\nonumber\\
i\,dM^I\equiv    \mathcal{U}^{2 I} &= \mathcal{U}^{2 I}_aV^a+ i\left( \widebar\uppsi +\widebar\uppsi^* \gamma_5\right)\chi^I = {\mathcal{U}}^{2 I}_aV^a - i  \widebar \chi^I \left( \uppsi + \gamma_5 \uppsi^* \right)\\
    &=\mathcal{U}^{2 I}_aV^a -\widebar\psi_1 \zeta_\Ibar  \delta^{I\Ibar}+\widebar\psi^2\zeta^I
\nonumber\,,
\\
    \widebar E^\Ibar \equiv   \;\mathcal{U}^{2 \Ibar}&= \mathcal{U}^{2 \Ibar}_aV^a+ \left( \widebar\uppsi +\widebar\uppsi^* \gamma_5\right)\chi^\Ibar= \mathcal{U}^{2 \Ibar}_aV^a+  \widebar\chi^{\Ibar}\left(  \uppsi + \gamma_5\uppsi^* \right)
    \label{ebar}\nonumber\\
    &= \mathcal{U}^{2 \Ibar}_a V^a + \widebar\psi_1\zeta_I \delta^{I\Ibar}+ \widebar\psi^2\zeta^\Ibar \,,
   \\ 
 i \,d\widebar M^\Ibar\equiv   \mathcal{U}^{1 \Ibar}&= \mathcal{U}^{1 \Ibar}_aV^a- i\left( \widebar\uppsi -\widebar\uppsi^* \gamma_5\right)\chi^\Ibar= \mathcal{U}^{1 \Ibar}_aV^a+ i\widebar\chi^{\Ibar}\left(  \uppsi - \gamma_5\uppsi^* \right)\nonumber\\
    &= \mathcal{U}^{1 \Ibar}_aV^a + \widebar\psi^1\zeta^\Ibar-\widebar\psi_2\zeta_I \delta^{I\Ibar} \,.
\end{align}
\end{subequations}
Here, the Dirac spinors  $\chi^I, \chi^\Ibar$ and the  adjoint spinors $\widebar\chi_\Ibar, \widebar\chi_I$ are related to the left-handed and right-handed projections of the hypermultiplet spinors $\zeta_\upalpha,\zeta^\upalpha$ in the following way:
\begin{subequations}\begin{align}
    \chi^I \equiv  \left( \zeta^I+i\zeta_\Jbar \delta^{I\Jbar} \right) \,,\quad
    & \widebar\chi^I \equiv  \left(\bar\zeta^I -i\,\bar\zeta_\Ibar \delta^{I\Ibar} \right) \,,\\
    \chi^\Ibar =   \left( \zeta_I\delta^{I\Ibar}+i\zeta^\Ibar \right) \,,\quad
    & \widebar\chi^\Ibar \equiv  \left(\bar\zeta_J \delta^{J\Ibar}
    - i\bar\zeta^\Ibar \right).
\end{align}\end{subequations}

At this point, we find convenient to introduce the following Dirac spinor 1-form combinations (see also Appendix \ref{conventions}):
\begin{equation}
\nonumber
    \begin{split}
        \uppsi_\pm &\equiv \frac 12 \left(\uppsi \pm \gamma_5\uppsi^*\right) \,, \quad \left\{ \begin{matrix}
           \uppsi_+ = \psi_1 + i \psi^2\\
           \uppsi_- = \psi^1 + i \psi_2
        \end{matrix} \,,
        \right.
    \end{split}
\end{equation}
and their adjoint,
\begin{equation}
\nonumber
    \begin{split}
        \widebar\uppsi_\pm &\equiv \frac 12 \left(\widebar\uppsi \pm \widebar\uppsi^*\gamma_5\right)\,, \quad \left\{ \begin{matrix}
           \widebar\uppsi_+ = \widebar\psi_1 - i \widebar\psi^2\\
           \widebar\uppsi_- = \widebar\psi^1 - i \widebar\psi_2
        \end{matrix} \,.
        \right.
    \end{split}
\end{equation}
In terms of them, eqs. \eqref{hypscapm} take the simpler form:
\begin{subequations}\label{hypscapm2}
\begin{align}
    E^I &\equiv\;\mathcal{U}^{1 I} = \mathcal{U}^{1 I}_aV^a+ \widebar\uppsi_-\chi^I 
    \label{e2}\\
  dM^I &\equiv -i\,  \mathcal{U}^{2 I} = -i\,\mathcal{U}^{2 I}_aV^a+   \widebar\uppsi_+\chi^I \,,
\\
    \widebar E^\Ibar &\equiv   \;\mathcal{U}^{2 \Ibar} = \mathcal{U}^{2 \Ibar}_aV^a+ \widebar\uppsi_+\chi^\Ibar \,,
   \\ 
  d\widebar M^\Ibar&\equiv  -i\,\mathcal{U}^{1 \Ibar} = -i\,\mathcal{U}^{1 \Ibar}_aV^a- \widebar\uppsi_-\chi^\Ibar\,.
\end{align}
\end{subequations}
Correspondingly, the definitions and superspace parametrizations of the superfields read
\begin{subequations}\begin{align}
    E^I &\equiv E^I_i dz^i  \nonumber \\
    &= E^I_aV^a+ \widebar\uppsi_-\chi^I \,, \\
    H^I &\equiv dB^I +4\,i\,\widebar\uppsi_+\gamma_a\gamma_5\left(L^I(z) \uppsi_+ +M^I\uppsi_-
    \right)V^a\nonumber\\
    &= H^I_{ abc}V^aV^bV^c-2\widebar\uppsi_+\gamma_5 \gamma_{ab}\chi^I V^aV^b\,, \\
    \mathcal{D} \chi^I&\equiv 
    \mathcal{D}_a \chi^I V^a + i\,E^I_a \gamma^a\uppsi_+ +\frac 32h^I_a\gamma^a\uppsi_- \,, 
\end{align}\end{subequations}
and
\begin{subequations}\begin{align}
    \widebar E^\Ibar &\equiv \widebar E^\Ibar_i dz^i  \nonumber \\
    &=\widebar E^\Ibar_aV^a+ \widebar\uppsi_+ \chi^\Ibar \,, \\
    \widebar H^\Ibar &\equiv d\widebar B^\Ibar +4\,i\,\widebar\uppsi_-\gamma_a\gamma_5\left(\widebar L^\Ibar(z) \uppsi_- - \widebar M^\Ibar\uppsi_+
    \right)V^a\nonumber\\
    &= \widebar H^\Ibar_{ abc}V^aV^bV^c-2\widebar\uppsi_-\gamma_5 \gamma_{ab}\chi^\Ibar V^aV^b\,, \\
    \mathcal{D} \chi^\Ibar &\equiv 
    \mathcal{D}_a \chi^\Ibar V^a + i\,\widebar E^\Ibar_a \gamma^a\uppsi_- +\frac 32\bar h^\Ibar_a\gamma^a\uppsi_+ \,.
\end{align}\end{subequations}
The formulation of the $\mathcal{N}=2$ model in terms of one Dirac gravitino makes it evident that the complex structure of the supersymmetric background, which is associated with the $U(1)\subset SU(2)_R$: $\uppsi \to \uppsi^*$, is different from the one acting on the double-tensor multiplets, which is instead:  $\uppsi_\pm \to \uppsi_\mp$.

\subsubsection*{The double-tensor multiplets Lagrangian with Dirac spinors}

In terms of the Dirac spinor quantities above, the $\mathcal{N}=2$ superspace Lagrangian of the double-tensor multiplets \eqref{superlag} boils down to
\begin{align}  
    \mathcal{L}_{\text{superspace}}& = a_1 \delta_{I\Jbar}\Bigl\{\left[\tilde{E}^{I|a}\left(\widebar E^\Jbar- \widebar\uppsi_+\chi^\Jbar\right) + \tilde{\widebar E}^{\Jbar|a}\left( E^I- \widebar\uppsi_-\chi^I \right)\right]V^bV^cV^d\epsilon_{abcd} \nonumber \\
    &-\frac 94\left[{\tilde{\bar h}^\Jbar_c }\left(H^I -i\widebar\uppsi_+\gamma_{ab}\gamma_5\chi^I
    V^aV^b\right) + {\tilde h^J_c }\left(\widebar H^\Jbar +i\widebar\uppsi_-\gamma_{ab}\gamma_5\chi^\Ibar      V^aV^b\right)\right]V^c\nonumber \\ 
    &-\frac 14 \left(\tilde{E}^{I|\ell}\tilde{\widebar{E}}^{\Jbar }_\ell -\frac 94 {\tilde{h}^{I|\ell}\tilde{\bar h}^{\Jbar }_\ell } \right)V^aV^bV^cV^d\epsilon_{abcd}
    \nonumber \\
    &- \frac{i}{2} \left(\ \widebar\chi^I\gamma^a\mathcal{D}\chi^\Jbar  +\widebar\chi^\Jbar\gamma^a\mathcal{D}\chi^I \right)V^bV^cV^d\epsilon_{abcd} \nonumber \\
    &-3i \left[E^I \widebar\uppsi_+\gamma_{ab}\gamma_5\chi^\Jbar 
   + \widebar E^\Jbar \widebar\uppsi_-\gamma_{ab}\gamma_5\chi^I \right]  V^aV^b \nonumber \\
    & - \frac 32\,i\, \left(H^I \widebar\uppsi_-\chi^\Jbar +\widebar H^\Jbar \widebar\uppsi_+\chi^I\right)  \nonumber \\
    &+ 3i\,\widebar\chi^\Jbar\gamma_a \chi^I\widebar\uppsi_-\gamma_b\gamma_5\uppsi_+V^a V^b\nonumber\\
    &- \frac 32 i \left(\widebar\chi^I\chi^\Jbar \widebar\uppsi_-\gamma_{ab}\gamma_5\uppsi_++\widebar\chi^I\gamma_5\chi^\Jbar \widebar\uppsi_-\gamma_{ab}\uppsi_+\right)V^aV^b\nonumber\\
    &+3 \left(L^I \widebar E^{\Jbar}-   \widebar L^\Jbar E^I \right)\widebar\uppsi_-\gamma_a\gamma_5\uppsi_+V^a \nonumber\\
    &-3i \left( M^I \mathcal{U}^{1\Jbar} - \widebar M^\Jbar \mathcal{U}^{2I} \right)  \widebar\uppsi_-\gamma_a\gamma_5\uppsi_+V^a\Bigr\} \,.
\end{align}
The relations found for the $\mathcal{N}=2$ double-tensor multiplets disclose an important difference between the $\mathcal{N}=1$ case: In the double-tensor model, the consistency of the theory, namely the closure of the Bianchi identities, requires the
scalars $M^I,\widebar M^\Ibar$, Hodge-dually related to the 2-forms, to appear naked in the superspace extension of the tensor field strengths, as we can see from \eqref{defH} and \eqref{relhq}. This is not the case for the $\mathcal{N}=1$, where the scalars of the theory prior to dualization do not appear anymore after dualizing them into tensors. In the $\mathcal{N}=1$ theory, both the off-shell matching of bosonic and fermionic d.o.f. and the off-shell closure of the supersymmetry algebra are realized.

\end{document}